\documentclass[10pt]{article}
\usepackage{graphicx}
\usepackage{amsmath}
\usepackage{amssymb}
\usepackage{caption2}
\begin{document}
\bibliographystyle{prsty}
\begin{center}
{\large {\bf \sc{  Analysis of  the ${\frac{1}{2}}^{\pm}$  pentaquark states in the diquark model with QCD sum rules }}} \\[2mm]
Zhi-Gang  Wang$^{1}$ \footnote{E-mail: zgwang@aliyun.com.  }, Tao Huang$^{2}$ \footnote{Email: huangtao@ihep.ac.cn.}     \\
$^{1}$ Department of Physics, North China Electric Power University, Baoding 071003, P. R. China \\
$^{2}$ Institute of High Energy Physics and Theoretical Physics
Center for Science Facilities, Chinese Academy of Sciences, Beijing 100049, P.R. China
\end{center}

\begin{abstract}
In this article, we present the scalar-diquark-scalar-diquark-antiquark type and scalar-diquark-axialvector-diquark-antiquark type pentaquark configurations in the diquark model, and  study the masses and pole residues of the $J^P={\frac{1}{2}}^\pm$ hidden-charm pentaquark states in details with the QCD sum rules by extending   our previous  work  on the $J^P={\frac{3}{2}}^-$ and ${\frac{5}{2}}^{+}$ hidden-charm pentaquark states. We calculate the contributions of the vacuum condensates up to dimension-10 in the operator product expansion by constructing both the scalar-diquark-scalar-diquark-antiquark type and scalar-diquark-axialvector-diquark-antiquark type interpolating currents. The present predictions of the masses can be confronted to the LHCb experimental data in the future.
\end{abstract}



\section{Introduction}

Recently,  the  LHCb collaboration  observed  two pentaquark candidates $P_c(4380)$ and $P_c(4450)$ in the $J/\psi p$ mass spectrum in the $\Lambda_b^0\to J/\psi K^- p$ decays with the significances of more than $9\,\sigma$  \cite{LHCb-4380}. The  measured masses and widths are  $M_{P_c(4380)}=4380\pm 8\pm 29\,\rm{MeV}$, $M_{P_c(4450)}=4449.8\pm 1.7\pm 2.5\,\rm{MeV}$, $\Gamma_{P_c(4380)}=205\pm 18\pm 86\,\rm{MeV}$ and  $\Gamma_{P_c(4450)}=39\pm 5\pm 19\,\rm{MeV}$, respectively. The $P_c(4380)$ and $P_c(4450)$  have the preferred spin-parity     $J^P={\frac{3}{2}}^-$ and ${\frac{5}{2}}^+$, respectively.
 The decays $P_c(4380)\to J/\psi p$ take place through relative S-wave while the decays $P_c(4450)\to J/\psi p$ take place through relative P-wave, the decays  $P_c(4450)\to J/\psi p$ are suppressed in the phase space, so the $P_c(4450)$ has smaller width.
 There have been  several  attempted assignments, such as the $\Sigma_c \bar{D}^*$, $\Sigma_c^* \bar{D}^*$, $\chi_{c1}p$, $J/\psi N(1440)$, $J/\psi N(1520)$ molecule-like  pentaquark  states \cite{mole-penta} (or not the molecular pentaquark states \cite{mole-penta-No}),  the diquark-diquark-antiquark type pentaquark states \cite{di-di-anti-penta,Wang1508}, the diquark-triquark  type  pentaquark states \cite{di-tri-penta}, re-scattering effects \cite{rescattering-penta}, etc. We can test their resonant nature by using photoproduction off a proton target \cite{Test-Penta}.

In Ref.\cite{Wang1508}, we construct the scalar-diquark-axialvector-diquark-antiquark type interpolating currents, calculate the contributions of the vacuum condensates up to dimension-10 in the operator product expansion, and  extend the energy scale formula suggested in our previous works \cite{Wang-Tetraquark-DW} to study the masses and pole residues  of the $J^P={\frac{3}{2}}^-$ and ${\frac{5}{2}}^+$ hidden-charm pentaquark states    with the QCD sum rules, and   assign the $P_c(4380)$ and $P_c(4450)$ to be the ${\frac{3}{2}}^-$ and ${\frac{5}{2}}^+$ pentaquark states, respectively.
 In this article, we extend our previous work to study the   $J^P={\frac{1}{2}}^{\pm}$  diquark-diquark-antiquark type  hidden charm pentaquark state by calculating the contributions of the vacuum condensates up to dimension-10, and try to obtain the lowest masses based on the QCD sum rules.

 The article is arranged as follows:  we choose the optimal pentaquark configurations in Sect.2; in Sect.3,
  we derive the QCD sum rules for the masses and pole residues of  the
$ {\frac{1}{2}}^{\pm}$ pentaquark states;  in Sect.4, we present the numerical results; and Sect.5 is reserved for our
summary   and discussions.

\section{Pentaquark configurations in the diquark model}
 The diquarks $q^{T}_j C\Gamma q^{\prime}_k$ have  five  structures  in Dirac spinor space, where $C\Gamma=C\gamma_5$, $C$, $C\gamma_\mu \gamma_5$,  $C\gamma_\mu $ and $C\sigma_{\mu\nu}$ for the scalar, pseudoscalar, vector, axialvector  and  tensor diquarks, respectively, and the $j$ and $k$ are color indexes. The matrices
$C\gamma_\mu $ and $C\sigma_{\mu\nu}$ are symmetric, the matrices
$C\gamma_5$, $C$ and $C\gamma_\mu \gamma_5$ are antisymmetric. The
attractive interactions of one-gluon exchange  favor  formation of
the diquarks in  color antitriplet $\overline{3}_{ c}$, flavor
antitriplet $\overline{3}_{ f}$ and spin singlet $1_s$ \cite{One-gluon},
 while the favored configurations are the scalar diquark states ($\varepsilon^{ijk}q^{T}_j C\gamma_5 q^{\prime}_k$) and axialvector diquark states ($\varepsilon^{ijk}q^{T}_j C\gamma_\mu q^{\prime}_k$) \cite{WangDiquark,WangLDiquark}.  The calculations based on the QCD sum rules indicate that the heavy-light scalar and axialvector diquark states have almost  degenerate masses \cite{WangDiquark},  while the masses of the  light axialvector  diquark states lie  about  $(150-200)\,\rm{MeV}$ above that of the light  scalar  diquark states    \cite{WangLDiquark}, if they have the same quark constituents.
  In this article, we take the diquark states as  basic constituents, and choose the scalar-diquark-scalar-diquark-antiquark type and scalar-diquark-axialvector-diquark-antiquark type pentaquark configurations.

  Now we illustrate how to construct the pentaquark states in the diquark model  according to the spin-parity $J^P$,
\begin{eqnarray}
0^+_{ud}\otimes 0^+_{uc} \otimes {\frac{1}{2}}^-_{\bar{c}} &=& \underline{{\frac{1}{2}}^-_{uudc\bar{c}}} \, ,  \\
0^+_{ud}\otimes 1^+_{uc} \otimes {\frac{1}{2}}^-_{\bar{c}} &=& \underline{{\frac{1}{2}}^-_{uudc\bar{c}}} \oplus \overline{{\frac{3}{2}}^-_{uudc\bar{c}}} \, , \\
0^+_{ud}\otimes 0^+_{uc} \otimes \left[1^-\otimes {\frac{1}{2}}^-_{\bar{c}} \right]&=& 0^+_{ud}\otimes 0^+_{uc} \otimes \left[ {\frac{1}{2}}^+_{\bar{c}} \oplus{\frac{3}{2}}^+_{\bar{c}} \right]= \underline{{\frac{1}{2}}^+_{uudc\bar{c}}}\oplus {\frac{3}{2}}^+_{uudc\bar{c}}\, ,  \\
0^+_{ud}\otimes 1^+_{uc} \otimes \left[1^-\otimes {\frac{1}{2}}^-_{\bar{c}} \right]&=& 0^+_{ud}\otimes 1^+_{uc} \otimes \left[ {\frac{1}{2}}^+_{\bar{c}} \oplus{\frac{3}{2}}^+_{\bar{c}} \right]\nonumber\\
&=& \left[\underline{{\frac{1}{2}}^+_{uudc\bar{c}}}\oplus {\frac{3}{2}}^+_{uudc\bar{c}}\right]\oplus \left[ {\frac{1}{2}}^+_{uudc\bar{c}}\oplus {\frac{3}{2}}^+_{uudc\bar{c}} \oplus \overline{{\frac{5}{2}}^+_{uudc\bar{c}}} \right] \, ,
\end{eqnarray}
where the $1^-$ denotes the contribution of the additional P-wave to the spin-parity, the subscripts $ud$, $uc$, $\bar{c}$ and $uudc\bar{c}$ denote the quark constituents. The quark and antiquark have opposite parity,   we usually take it for granted that the quarks have positive parity while the antiquarks have negative parity, so the $\bar{c}$-quark has $J^P={\frac{1}{2}}^-$.

The overlined states  ${\frac{3}{2}}^-_{uudc\bar{c}}$ and ${\frac{5}{2}}^+_{uudc\bar{c}}$ are assigned to be the pentaquark states $P_c(4380)$ and $P_c(4450)$,  respectively \cite{Wang1508}.
In previous work \cite{Wang1508}, we choose the scalar-diquark-axialvector-diquark-antiquark type currents  $J_\mu(x)$ and $J_{\mu\nu}(x)$,
\begin{eqnarray}
 J_{\mu}(x)&=&\varepsilon^{ila} \varepsilon^{ijk}\varepsilon^{lmn}  u^T_j(x) C\gamma_5 d_k(x)\,u^T_m(x) C\gamma_\mu c_n(x)\, C\bar{c}^{T}_{a}(x) \, , \\
 J_{\mu\nu}(x)&=&\frac{1}{\sqrt{2}}\varepsilon^{ila} \varepsilon^{ijk}\varepsilon^{lmn}  u^T_j(x) C\gamma_5 d_k(x)\left[u^T_m(x) C\gamma_\mu c_n(x)\, \gamma_{\nu}C\bar{c}^{T}_{a}(x)+u^T_m(x) C\gamma_\nu c_n(x)\, \gamma_{\mu}C\bar{c}^{T}_{a}(x)\right] \, ,\nonumber\\
\end{eqnarray}
to interpolate the ${\frac{3}{2}}^-$ and ${\frac{5}{2}}^+$ pentaquark states, respectively, where the $i$, $j$, $k$, $\cdots$ are color indices, the $C$ is the charge conjugation matrix.

   The underlined states ${\frac{1}{2}}^-_{uudc\bar{c}}$  are supposed to be the lowest pentaquark states, while their P-wave partners  ${\frac{1}{2}}^+_{uudc\bar{c}}$  are supposed to be the lowest pentaquark states with the positive parity. In this article, we   choose   both the scalar-diquark-scalar-diquark-antiquark type and scalar-diquark-axialvector-diquark-antiquark type currents $J_{j_L j_H}(x)$,
    \begin{eqnarray}
 J_{00}(x)&=&\varepsilon^{ila} \varepsilon^{ijk}\varepsilon^{lmn}  u^T_j(x) C\gamma_5 d_k(x)\,u^T_m(x) C\gamma_5 c_n(x)\, \gamma_5 C\bar{c}^{T}_{a}(x) \, , \\
  J_{01}(x)&=&\varepsilon^{ila} \varepsilon^{ijk}\varepsilon^{lmn}  u^T_j(x) C\gamma_5 d_k(x)\,u^T_m(x) C\gamma_\mu c_n(x)\, \gamma^\mu C\bar{c}^{T}_{a}(x) \, ,
\end{eqnarray}
    to study the lowest pentaquark states with $J^P={\frac{1}{2}}^{\pm}$ in a consistent way, where the subscripts  $j_L$ and $j_H$ denote the spins of the light and heavy diquarks, respectively.

\section{QCD sum rules for  the $ {\frac{1}{2}}^{\pm}$ pentaquark states}

In the following, we write down  the two-point correlation functions $\Pi_{j_L j_H}(p)$  in the QCD sum rules,
\begin{eqnarray}
\Pi_{j_L j_H}(p)&=&i\int d^4x e^{ip \cdot x} \langle0|T\left\{J_{j_L j_H}(x)\bar{J}_{j_L j_H}(0)\right\}|0\rangle \, .
\end{eqnarray}
The currents $J_{j_L j_H}(0)$ have positive parity, and    couple potentially to the ${\frac{1}{2}}^+$   hidden-charm  pentaquark  states $P_{j_L j_H}^{+}$,
\begin{eqnarray}
\langle 0| J_{j_L j_H} (0)|P_{j_L j_H}^{+}(p)\rangle &=&\lambda^{+}_{j_L j_H} U^{+}(p,s) \, ,
\end{eqnarray}
the $\lambda^+_{j_L j_H}$ are the pole residues, the spinors $U^{+}(p,s)$   satisfy the Dirac equations $(\not\!\!p-M_{j_L j_H,+})U^{+}(p)=0$.   On the other hand, the currents $J_{j_L j_H}(0)$  also couple potentially to the ${\frac{1}{2}}^-$   hidden-charm  pentaquark states $P_{j_L j_H}^{-}$ as multiplying $i \gamma_{5}$ to the currents $J_{j_L j_H}(x)$ changes their parity  \cite{Chung82,Bagan93,Oka96,WangHbaryon},
 \begin{eqnarray}
\langle 0| J_{j_L j_H} (0)|P_{j_L j_H}^{-}(p)\rangle &=&\lambda^{-}_{j_L j_H}i\gamma_5 U^{-}(p,s) \, ,
\end{eqnarray}
the spinors $U^{\pm}(p,s)$ (pole residues $\lambda_{j_L j_H}^\pm$)   have analogous  properties.

 We  insert  a complete set  of intermediate pentaquark states with the same quantum numbers as the current operators $J_{j_L j_H}(x)$,
  and $i\gamma_5 J_{j_L j_H}(x)$ into the correlation functions $\Pi_{j_L j_H}(p)$   to obtain the hadronic representation
\cite{SVZ79,PRT85}. After isolating the pole terms of the lowest states of the hidden-charm  pentaquark states, we obtain the
following results:
\begin{eqnarray}
  \Pi_{j_L j_H}(p) & = & {\lambda^{+}_{j_L j_H}}^2 \,\, {\!\not\!{p}+ M_{j_L j_H,+} \over M_{j_L j_H,+}^{2}-p^{2}  }  +   {\lambda^{-}_{j_L j_H}}^2\,\,  {\!\not\!{p}- M_{j_L j_H,-} \over M_{j_L j_H,-}^{2}-p^{2}  }     +\cdots \, ,
    \end{eqnarray}
where the $M_{j_L j_H,\pm}$ are the masses of the lowest pentaquark states with the
 parity $\pm$ respectively. We have to include the negative parity pentaquark states as $M_{j_L j_H,+}>M_{j_L j_H,-}$ according to the special quark configurations, see Eqs.(1-4).

Now we obtain the hadronic spectral densities  through the dispersion relation,
\begin{eqnarray}
\frac{{\rm Im}\Pi_{j_L j_H}(s)}{\pi}&=&\!\not\!{p} \left[{\lambda^{+}_{j_L j_H}}^2 \,\,\delta\left(s-M_{j_L j_H,+}^2\right)+{\lambda^{-}_{j_L j_H}}^2\,\, \delta\left(s-M_{j_L j_H,-}^2\right)\right] \nonumber\\
&& +\left[M_{j_L j_H,+}{\lambda^{+}_{j_L j_H}}^2 \,\,\delta\left(s-M_{j_L j_H,+}^2\right)-M_{j_L j_H,-}{\lambda^{-}_{j_L j_H}}^2 \,\,\delta\left(s-M_{j_L j_H,-}^2\right)\right]\, , \nonumber\\
&=&\!\not\!{p} \,\rho^1_{j_L j_H}(s)+\rho^0_{j_L j_H}(s) \, ,
\end{eqnarray}
then we introduce the weight function $\exp\left(-\frac{s}{T^2}\right)$ to obtain the QCD sum rules at the hadron  side,
\begin{eqnarray}
\int_{4m_c^2}^{s_0}ds \left[\sqrt{s}\rho^1_{j_L j_H}(s)+\rho^0_{j_L j_H}(s)\right]\exp\left( -\frac{s}{T^2}\right)
&=&2M_{j_L j_H,+}{\lambda^{+}_{j_L j_H}}^2\,\,\exp\left( -\frac{M_{j_L j_H,+}^2}{T^2}\right) \, , \\
\int_{4m_c^2}^{s_0}ds \left[\sqrt{s}\rho^1_{j_L j_H}(s)-\rho^0_{j_L j_H}(s)\right]\exp\left( -\frac{s}{T^2}\right)
&=&2M_{j_L j_H,-}{\lambda^{-}_{j_L j_H}}^2\,\,\exp\left( -\frac{M_{j_L j_H,-}^2}{T^2}\right) \, ,
\end{eqnarray}
where the $s_0$ are the continuum threshold parameters and the $T^2$ are the Borel parameters. We separate the contributions of   the
negative-parity (positive-parity) pentaquark states from the positive-parity (negative-parity) pentaquark states explicitly.

In the following we briefly outline  the operator product expansion for the correlation functions $\Pi_{j_L j_H}(p)$   in perturbative QCD. Firstly,  we contract the $u$, $d$ and $c$ quark fields in the correlation functions
$\Pi_{j_L j_H}(p)$    with Wick theorem, and obtain the results:
\begin{eqnarray}
\Pi_{00}(p)&=&i\,\varepsilon^{ila}\varepsilon^{ijk}\varepsilon^{lmn}\varepsilon^{i^{\prime}l^{\prime}a^{\prime}}\varepsilon^{i^{\prime}j^{\prime}k^{\prime}}
\varepsilon^{l^{\prime}m^{\prime}n^{\prime}}\int d^4x e^{ip\cdot x} \nonumber\\
&&\left\{   Tr\left[\gamma_5 D_{kk^\prime}(x) \gamma_5 C U^{T}_{jj^\prime}(x)C\right] \,Tr\left[\gamma_5 C_{nn^\prime}(x) \gamma_5 C U^{T}_{mm^\prime}(x)C\right] \gamma_5 C C_{a^{\prime}a}^T(-x)C \gamma_5 \right. \nonumber\\
&&\left.-  Tr \left[\gamma_5 D_{kk^\prime}(x) \gamma_5 C U^{T}_{mj^\prime}(x)C \gamma_5 C_{nn^\prime}(x) \gamma_5 C U^{T}_{jm^\prime}(x)C\right]  \gamma_5 C C_{a^{\prime}a}^T(-x)C \gamma_5  \right\} \, ,
\end{eqnarray}
\begin{eqnarray}
\Pi_{01}(p)&=&i\,\varepsilon^{ila}\varepsilon^{ijk}\varepsilon^{lmn}\varepsilon^{i^{\prime}l^{\prime}a^{\prime}}\varepsilon^{i^{\prime}j^{\prime}k^{\prime}}
\varepsilon^{l^{\prime}m^{\prime}n^{\prime}}\int d^4x e^{ip\cdot x} \nonumber\\
&&\left\{   Tr\left[\gamma_5 D_{kk^\prime}(x) \gamma_5 C U^{T}_{jj^\prime}(x)C\right] \,Tr\left[\gamma_\mu C_{nn^\prime}(x) \gamma_\nu C U^{T}_{mm^\prime}(x)C\right]\gamma^{\mu} C C_{a^{\prime}a}^T(-x)C\gamma^{\nu} \right. \nonumber\\
&&\left.-  Tr \left[\gamma_5 D_{kk^\prime}(x) \gamma_5 C U^{T}_{mj^\prime}(x)C \gamma_\mu C_{nn^\prime}(x) \gamma_\nu C U^{T}_{jm^\prime}(x)C\right] \gamma^\mu C C_{a^{\prime}a}^T(-x)C \gamma^\nu \right\} \, ,
\end{eqnarray}
where
the $U_{ij}(x)$, $D_{ij}(x)$ and $C_{ij}(x)$ are the full $u$, $d$ and $c$ quark propagators respectively ($S_{ij}(x)=U_{ij}(x),\,D_{ij}(x)$),
 \begin{eqnarray}
S_{ij}(x)&=& \frac{i\delta_{ij}\!\not\!{x}}{ 2\pi^2x^4}-\frac{\delta_{ij}\langle
\bar{q}q\rangle}{12} -\frac{\delta_{ij}x^2\langle \bar{q}g_s\sigma Gq\rangle}{192} -\frac{ig_sG^{a}_{\alpha\beta}t^a_{ij}(\!\not\!{x}
\sigma^{\alpha\beta}+\sigma^{\alpha\beta} \!\not\!{x})}{32\pi^2x^2} \nonumber\\
&&  -\frac{1}{8}\langle\bar{q}_j\sigma^{\mu\nu}q_i \rangle \sigma_{\mu\nu}+\cdots \, ,
\end{eqnarray}
\begin{eqnarray}
C_{ij}(x)&=&\frac{i}{(2\pi)^4}\int d^4k e^{-ik \cdot x} \left\{
\frac{\delta_{ij}}{\!\not\!{k}-m_c}
-\frac{g_sG^n_{\alpha\beta}t^n_{ij}}{4}\frac{\sigma^{\alpha\beta}(\!\not\!{k}+m_c)+(\!\not\!{k}+m_c)
\sigma^{\alpha\beta}}{(k^2-m_c^2)^2}\right.\nonumber\\
&&\left. -\frac{g_s^2 (t^at^b)_{ij} G^a_{\alpha\beta}G^b_{\mu\nu}(f^{\alpha\beta\mu\nu}+f^{\alpha\mu\beta\nu}+f^{\alpha\mu\nu\beta}) }{4(k^2-m_c^2)^5}+\cdots\right\} \, ,\nonumber\\
f^{\alpha\beta\mu\nu}&=&(\!\not\!{k}+m_c)\gamma^\alpha(\!\not\!{k}+m_c)\gamma^\beta(\!\not\!{k}+m_c)\gamma^\mu(\!\not\!{k}+m_c)\gamma^\nu(\!\not\!{k}+m_c)\, ,
\end{eqnarray}
and  $t^n=\frac{\lambda^n}{2}$, the $\lambda^n$ is the Gell-Mann matrix   \cite{PRT85}, then compute  the integrals both in the coordinate and momentum spaces to obtain the correlation functions $\Pi_{j_L j_H}(p)$,   therefore the QCD spectral densities $\rho_{j_L j_H,QCD}^1(s)$ and $\rho_{j_L j_H,QCD}^0(s)$ at the quark level through the dispersion  relation,
 \begin{eqnarray}
\frac{{\rm Im}\Pi_{j_L j_H}(s)}{\pi}&=&\!\not\!{p} \,\rho^1_{j_L j_H,QCD}(s)+\rho^0_{j_L j_H,QCD}(s) \, .
\end{eqnarray}
 In Eq.(18), we retain the term $\langle\bar{q}_j\sigma_{\mu\nu}q_i \rangle$  comes from the Fierz re-arrangement of the $\langle q_i \bar{q}_j\rangle$ to  absorb the gluons  emitted from other  quark lines to form $\langle\bar{q}_j g_s G^a_{\alpha\beta} t^a_{mn}\sigma_{\mu\nu} q_i \rangle$   so as to extract the mixed condensate  $\langle\bar{q}g_s\sigma G q\rangle$.

 Once the analytical QCD spectral densities $\rho_{j_L j_H,QCD}^1(s)$ and $\rho_{j_L j_H,QCD}^0(s)$ are obtained,  we can take the
quark-hadron duality below the continuum thresholds  $s_0$ and introduce the weight function $\exp\left(-\frac{s}{T^2}\right)$ to obtain  the following QCD sum rules:
\begin{eqnarray}
2M_{j_L j_H,+}{\lambda^{+}_{j_L j_H}}^2\,\,\exp\left( -\frac{M_{j_L j_H,+}^2}{T^2}\right)
&=& \int_{4m_c^2}^{s_0}ds \left[\sqrt{s}\rho_{j_L j_H,QCD}^1(s)+\rho_{j_L j_H,QCD}^0(s)\right]\exp\left( -\frac{s}{T^2}\right)\, ,\nonumber\\
\end{eqnarray}
\begin{eqnarray}
2M_{j_L j_H,-}{\lambda^{-}_{j_L j_H}}^2\,\,\exp\left( -\frac{M_{j_L j_H,-}^2}{T^2}\right)
&=& \int_{4m_c^2}^{s_0}ds \left[\sqrt{s}\rho_{j_L j_H,QCD}^1(s)-\rho_{j_L j_H,QCD}^0(s)\right]\exp\left( -\frac{s}{T^2}\right)\, ,\nonumber\\
\end{eqnarray}
where $\rho_{j_L j_H,QCD}^0(s)={\bf -}m_c \widetilde{\rho}_{j_L j_H,QCD}^0(s)$,
\begin{eqnarray}
\rho_{j_L j_H,QCD}^1(s)&=&\rho_{j_L j_H,0}^1(s)+\rho_{j_L j_H,3}^1(s)+\rho_{j_L j_H,4}^1(s)+\rho_{j_L j_H,5}^1(s)+\rho_{j_L j_H,6}^1(s)+\rho_{j_L j_H,8}^1(s)\nonumber\\
&&+\rho_{j_L j_H,9}^1(s)+\rho_{j_L j_H,10}^1(s)\, , \nonumber\\
\widetilde{\rho}_{j_L j_H,QCD}^0(s)&=&\widetilde{\rho}_{j_L j_H,0}^0(s)+\widetilde{\rho}_{j_L j_H,3}^0(s)+\widetilde{\rho}_{j_L j_H,4}^0(s)+\widetilde{\rho}_{j_L j_H,5}^0(s)+\widetilde{\rho}_{j_L j_H,6}^0(s)
+\widetilde{\rho}_{j_L j_H,8}^0(s)\nonumber\\
&&+\widetilde{\rho}_{j_L j_H,9}^0(s)+\widetilde{\rho}_{j_L j_H,10}^0(s)\, ,
\end{eqnarray}
the explicit expressions of the  QCD spectral densities $\rho_{j_L j_H,i}^1(s)$ and $\widetilde{\rho}_{j_L j_H,i}^0(s)$ with $i=0$, $3$, $4$, $5$, $6$, $8$, $9$, $10$ are shown in the appendix. Here we introduce a negative sign in the definition $\rho_{j_L j_H,QCD}^0(s)={\bf -}m_c \widetilde{\rho}_{j_L j_H,QCD}^0(s)$ to warrant positive spectral densities  $\widetilde{\rho}_{j_L j_H,QCD}^0(s)$,
\begin{eqnarray}
\int_{4m_c^2}^{s_0} ds\,\widetilde{\rho}_{j_L j_H,QCD}^0(s)\,\exp\left( -\frac{s}{T^2}\right) &> & 0\, .
\end{eqnarray}
In this article, we carry out the
operator product expansion to the vacuum condensates  up to dimension-10, and
assume vacuum saturation for the  high dimension vacuum condensates.

We differentiate   Eqs.(21-22) with respect to  $\frac{1}{T^2}$, then eliminate the
 pole residues $\lambda^{\pm}_{j_L j_H}$ and obtain the QCD sum rules for
 the masses of the pentaquark states,
 \begin{eqnarray}
 M^2_{j_L j_H,+} &=& \frac{\int_{4m_c^2}^{s_0}ds \left[\sqrt{s}\rho_{j_L j_H,QCD}^1(s)-m_c\widetilde{\rho}_{j_L j_H,QCD}^0(s)\right]\,s\,\exp\left( -\frac{s}{T^2}\right)}{\int_{4m_c^2}^{s_0}ds \left[\sqrt{s}\rho_{j_L j_H,QCD}^1(s)-m_c\widetilde{\rho}_{j_L j_H,QCD}^0(s)\right]\exp\left( -\frac{s}{T^2}\right)}\, ,\\
M^2_{j_L j_H,-} &=&\frac{\int_{4m_c^2}^{s_0}ds \left[\sqrt{s}\rho_{j_L j_H,QCD}^1(s)+m_c\widetilde{\rho}_{j_L j_H,QCD}^0(s)\right]\,s\,\exp\left( -\frac{s}{T^2}\right)}{\int_{4m_c^2}^{s_0}ds \left[\sqrt{s}\rho_{j_L j_H,QCD}^1(s)+m_c\widetilde{\rho}_{j_L j_H,QCD}^0(s)\right]\exp\left( -\frac{s}{T^2}\right)}\, .
\end{eqnarray}

\section{Numerical results}
We take the vacuum condensates to be  the standard values
$\langle\bar{q}q \rangle=-(0.24\pm 0.01\, \rm{GeV})^3$,
$\langle\bar{q}g_s\sigma G q \rangle=m_0^2\langle \bar{q}q \rangle$,
$m_0^2=(0.8 \pm 0.1)\,\rm{GeV}^2$, $\langle \frac{\alpha_s
GG}{\pi}\rangle=(0.33\,\rm{GeV})^4 $    at the energy scale  $\mu=1\, \rm{GeV}$
\cite{SVZ79,PRT85}.
The quark condensates and mixed quark condensates  evolve with the   renormalization group equation,
$\langle\bar{q}q \rangle(\mu)=\langle\bar{q}q \rangle(Q)\left[\frac{\alpha_{s}(Q)}{\alpha_{s}(\mu)}\right]^{\frac{4}{9}}$ and
 $\langle\bar{q}g_s \sigma Gq \rangle(\mu)=\langle\bar{q}g_s \sigma Gq \rangle(Q)\left[\frac{\alpha_{s}(Q)}{\alpha_{s}(\mu)}\right]^{\frac{2}{27}}$.
In the article, we take the $\overline{MS}$ mass $m_{c}(m_c)=(1.275\pm0.025)\,\rm{GeV}$
 from the Particle Data Group \cite{PDG}, and take into account
the energy-scale dependence of  the $\overline{MS}$ mass from the renormalization group equation,
\begin{eqnarray}
m_c(\mu)&=&m_c(m_c)\left[\frac{\alpha_{s}(\mu)}{\alpha_{s}(m_c)}\right]^{\frac{12}{25}} \, ,\nonumber\\
\alpha_s(\mu)&=&\frac{1}{b_0t}\left[1-\frac{b_1}{b_0^2}\frac{\log t}{t} +\frac{b_1^2(\log^2{t}-\log{t}-1)+b_0b_2}{b_0^4t^2}\right]\, ,
\end{eqnarray}
  where $t=\log \frac{\mu^2}{\Lambda^2}$, $b_0=\frac{33-2n_f}{12\pi}$, $b_1=\frac{153-19n_f}{24\pi^2}$, $b_2=\frac{2857-\frac{5033}{9}n_f+\frac{325}{27}n_f^2}{128\pi^3}$,  $\Lambda=213\,\rm{MeV}$, $296\,\rm{MeV}$  and  $339\,\rm{MeV}$ for the flavors  $n_f=5$, $4$ and $3$, respectively  \cite{PDG}.

In this article, we study the pentaquark configurations  consist of a light-diquark, a charm  diquark, a charm antiquark,  and
resort to the diquark-diquark-antiquark model to construct the currents to interpolate the hidden-charm pentaquark states.
The hidden charm (or bottom) five-quark systems  $qq_1q_2Q\bar{Q}$ could be described
by a double-well potential.   In the five-quark system $qq_1q_2Q\bar{Q}$, the light quarks $q_1$ and $q_2$ combine together to form a light diquark $\mathcal{D}_{q_1q_2}^j$ in  color antitriplet,
\begin{eqnarray}
q_1+q_2 &\to & \mathcal{D}_{q_1q_2}^j \, ,
\end{eqnarray}
the $\bar{Q}$-quark serves  as a static well potential, which binds the light diquark $\mathcal{D}_{q_1q_2}^j$  to
form a heavy triquark $\mathcal{T}^i_{q_1q_2\bar{Q}}$ in color triplet,
\begin{eqnarray}
\mathcal{D}_{q_1q_2}^j+\bar{Q}^k &\to & \mathcal{T}^i_{q_1q_2\bar{Q}}\, ,
\end{eqnarray}
while  the $Q$-quark serves as another static well potential,
which binds the light quark $q$  to form a heavy diquark in color antitriplet,
\begin{eqnarray}
q+Q &\to & \mathcal{D}^i_{qQ} \, ,
\end{eqnarray}
where the $i$, $j$ and $k$ are color indexes.
Then the heavy diquark $\mathcal{D}^i_{qQ}$ in color antitriplet   combines the heavy triquark $\mathcal{T}^i_{q_1q_2\bar{Q}}$ in color triplet  to form a pentaquark state in color singlet.

Such  a doubly-heavy pentaquark state  is characterized by the effective heavy quark masses ${\mathbb{M}}_Q$ (or constituent
quark masses) and the virtuality $V=\sqrt{M^2_{P}-(2{\mathbb{M}}_Q)^2}$ (or bound energy not as robust), just like   the doubly-heavy four-quark states \cite{Wang-Tetraquark-DW,WangHuang-PRD,Wang-Tetraquark-DW-2, Wang-molecule,Wang-Octet}.   The  QCD sum rules have three typical energy scales $\mu^2$, $T^2$, $V^2$, we
 take the energy  scale, $ \mu^2=V^2={\mathcal{O}}(T^2)$, and obtain energy scale formula,
 \begin{eqnarray}
 \mu&=&\sqrt{M_{P}^2-(2{\mathbb{M}}_c)^2}\, ,
   \end{eqnarray}
   to determine the energy scales of the QCD spectral densities. In previous work \cite{Wang1508}, we take the  value ${\mathbb{M}}_c=1.8\,\rm{GeV}$ determined in the diquark-antidiquark type tetraquark states \cite{Wang-Tetraquark-DW,WangHuang-PRD}, and obtain the values $\mu=2.5\,\rm{GeV}$ and $\mu=2.6\,\rm{GeV}$ for the hidden charm pentaquark states $P_c(4380)$  and $P_c(4450)$, respectively. The energy scale formula works well.

In this article, we choose the  Borel parameters $T^2$ and continuum threshold
parameters $s_0$  to satisfy the  four  criteria:

$\bf{1_\cdot}$ Pole dominance at the phenomenological side;

$\bf{2_\cdot}$ Convergence of the operator product expansion;

$\bf{3_\cdot}$ Appearance of the Borel platforms;

$\bf{4_\cdot}$ Satisfying the energy scale formula.

It is difficult to satisfy  the criteria $\bf{1}$ and $\bf{2}$ in the QCD sum rules for the multiquark states.
In the QCD sum rules for the hidden charm (or bottom) tetraquark states (or pentaquark states), molecular states and molecule-like states, the integrals
 \begin{eqnarray}
 \int_{4m_Q^2}^{s_0} ds \rho_{QCD}(s)\exp\left(-\frac{s}{T^2} \right)\, ,
 \end{eqnarray}
are sensitive to the heavy quark masses $m_Q$, where the $\rho_{QCD}(s)$ denotes the QCD spectral densities.
Variations of the heavy quark masses lead to changes of integral ranges $ \int_{4m_Q^2}^{s_0}$ of the variable  $\bf{ds}$ besides the QCD spectral densities $\rho_{QCD}(s)$, therefore changes of the Borel windows and predicted masses and pole residues. In calculations, we observe that small variations of the heavy quark
masses $m_Q$ can lead to rather large changes of the predictions \cite{Wang-Tetraquark-DW,WangHuang-PRD,Wang-Tetraquark-DW-2, Wang-molecule,Wang-Octet},
some constraints are needed to specialize  the heavy quark masses $m_Q$.

Now we write down the definition for the pole contributions and use a toy-model spectral density to illustrate how to enhance  the pole contributions,
\begin{eqnarray}
 {\rm Pole} &=& \frac{ \int_{4m_c^2}^{s_0} ds \rho_{QCD}(s)\exp\left(-\frac{s}{T^2} \right)}{ \int_{4m_c^2}^{\infty} ds \rho_{QCD}(s)
 \exp\left(-\frac{s}{T^2} \right)} \, ,
 \end{eqnarray}
where
\begin{eqnarray}
\rho_{QCD}(s)&=&(s-4m_c^2)^k \, ,
\end{eqnarray}
with $k=0,\,1,\,2,\,3,\,4,\,5$. The simple spectral density $\rho_{QCD}(s)$ makes  sense, as we can simplify the calculation by taking the rough  approximations $\overline{m}_c^2=\frac{(y+z)m_c^2}{yz}\approx 4m_c^2$ and
$ \widetilde{m}_c^2=\frac{m_c^2}{y(1-y)}\approx 4m_c^2$, see the  QCD spectral densities in the appendix.
For the hidden-charm tetraquark states, $k\leq 4$; for the hidden-charm pentaquark states, $k\leq 5$.

In Fig.1, we plot the pole contribution with variations of the $c$-quark mass $m_c$ for the typical Borel parameter $T^2=3.5\,\rm{GeV}^2$ and  continuum threshold parameter $s_0=25\,\rm{GeV}^2$. From the figure, we can see that the pole contribution  decreases monotonously with the increase of the $m_c$ and $k$. The $\overline{MS}$ mass $m_c(m_c)=1.275\,\rm{GeV}$ at the energy scale $\mu=m_c$ cannot lead to pole contribution $\geq50\%$ for the hidden-charm pentaquark states as $k_{max}= 5$. A smaller $m_c(\mu)$ (or a larger energy scale $\mu$), for example, $m_c(\mu)=1.1\,\rm{GeV}$, can lead to the pole contribution $\geq50\%$. However, we cannot choose  large energy scales freely  to enhance the pole contribution, as the quark condensates and mixed condensates increase slowly but monotonously with the increase of energy scale, which slows  down the convergent speed  in the operator product expansion. In this article,  we resort to the energy scale formula $\mu=\sqrt{M_{P}^2-(2{\mathbb{M}}_c)^2}$ with the  value ${\mathbb{M}}_c=1.8\,\rm{GeV}$ determined in the tetraquark states \cite{Wang-Tetraquark-DW}  to determine the energy scales of the QCD spectral densities, which works well in the QCD sum rules for the pentaquark candidates $P_c(4380)$ and $P_c(4450)$ \cite{Wang1508}.

\begin{figure}
 \centering
 \includegraphics[totalheight=8cm,width=10cm]{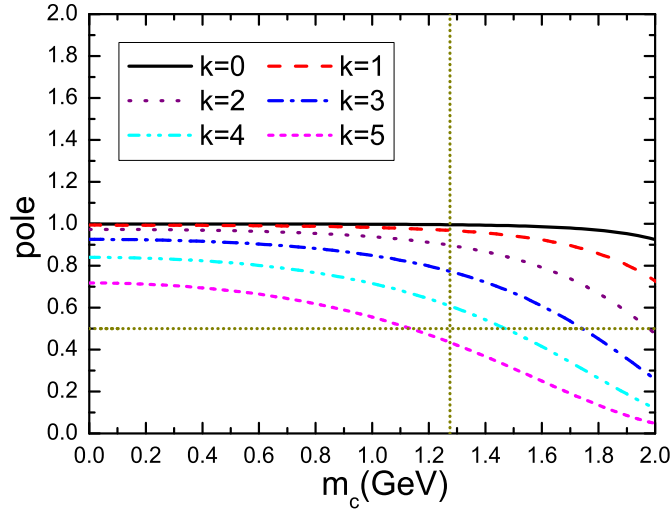}
        \caption{ The pole contributions with variations of the $m_c$ in the toy-model, where the perpendicular line corresponds to the $\overline{MS}$ mass $m_c(m_c)=1.275\,\rm{GeV}$.  }
\end{figure}

In previous work \cite{Wang-Tetraquark-DW,WangHuang-PRD}, we observed that the pole contributions can be taken as large as $(50-70)\%$ in the QCD sum rules for the diquark-antidiquark type tetraquark states $q\bar{q}^{\prime}Q\bar{Q}$ ($X,Y,Z$), if the QCD spectral densities obey  the energy scale formula $\mu=\sqrt{M_{X/Y/Z}^2-(2{\mathbb{M}}_Q)^2}$.  The operator product expansion converges
more slowly  in the QCD sum rules for the pentaquark states $qq_1q_2Q\bar{Q}$ compared  to  that for the  tetraquark states $q\bar{q}^{\prime}Q\bar{Q}$.
In Ref.\cite{Wang1508}, we observe that if we take  the energy scale formula to determine the QCD spectral densities, the pole contributions can reach  $(40-60)\%$. So in this article, we try to choose analogous   pole contributions,  $(50\pm10)\%$.

For the  tetraquark states $q\bar{q}^{\prime}Q\bar{Q}$ \cite{Wang-Tetraquark-DW,WangHuang-PRD}, the   Borel platforms appear as the minimum values, and the platforms are very flat, but the Borel windows are small, $T^2_{max}-T^2_{min}=0.4\,\rm{GeV}^2$, where the $max$ and $min$ denote the maximum and minimum values, respectively. For the heavy, doubly-heavy and triply-heavy baryon states $qq^{\prime}Q$, $qQQ^\prime$, $QQ^{\prime}Q^{\prime\prime}$  \cite{WangHbaryon,WangLambda},  the Borel platforms do not appear as the minimum values, the predicted masses increase slowly with the increase of the Borel parameter, we determine the Borel windows by the criteria $\bf{1}$ and $\bf{2}$, the platforms are not very flat. The pentaquark states are special baryon states, as they have one unit  baryon number. In this article, we also choose small Borel windows $T^2_{max}-T^2_{min}=0.4\,\rm{GeV}^2$, just like in the case of the tetraquark states \cite{Wang-Tetraquark-DW,WangHuang-PRD},  and obtain the platforms  by requiring the uncertainties $\frac{\delta M_{P}}{M_{P}} $ induced by the Borel parameters are about $1\%$.
In Ref.\cite{Wang1508}, we observe that such a criterion  can be  satisfied for the hidden-charm pentaquark states.

Now we  search for the optimal  Borel parameters $T^2$ and continuum threshold parameters $s_0$ according to  the four criteria. The resulting Borel parameters, continuum threshold parameters,  pole contributions, contributions of  the contributions of the vacuum condensates of dimension 9 and dimension 10  are shown explicitly in Table 1. From the Table, we can see that  the criteria $\bf{1}$ and $\bf{2}$ of the QCD sum rules are  satisfied.

 In calculations, we observe that
\begin{eqnarray}
\mu\uparrow  \, \, \, \, \,  M_{P} \downarrow \, ,\nonumber\\
\mu\downarrow  \, \, \, \, \,  M_{P} \uparrow \, ,
\end{eqnarray}
from the QCD sum rules in Eqs.(25-26). We can rewrite Eq.(31) into the following form,
\begin{eqnarray}
 M_{P}^2=4{\mathbb{M}}_c^2+\mu^2\, ,
   \end{eqnarray}
which indicates that
\begin{eqnarray}
\mu\uparrow  \, \, \, \, \,  M_{P} \uparrow \, ,\nonumber\\
\mu\downarrow  \, \, \, \, \, M_{P} \downarrow \, .
\end{eqnarray}
It is difficult  to obtain  the optimal energy scales $\mu$ and masses $M_{P}$, however, the optimal energy scales $\mu$ and masses $M_{P}$ do exist, see Table 2.

We take into account  all uncertainties  of the input   parameters,
and obtain the values of the masses and pole residues of
 the ${1\over 2}^{\pm}$   hidden-charm pentaquark states, which are shown in Figs.2-3 and Table 2. In Fig.2, we plot the masses with variations of the Borel parameters at large ranges, not just in the Borel windows. In the Borel windows, the uncertainties $\frac{\delta M_{P_c}}{M_{P_c}} $ induced by the Borel parameters  $\leq 1\%$.
 From Table 2, we can see that the predicted masses have the relations $M_{00,- }<M_{00,+} $ and $M_{01,- }<M_{01,+} $, which is consistent with our naive expectation, the pentaquark state with an additional P-wave has larger mass than corresponding S-wave state.     The value $M_{01,- }= 4.30\pm0.13 \,\rm{GeV}$ is smaller than the    value  $M_{P_c(4380) }=4.38\pm0.13\,\rm{GeV}$ \cite{Wang1508}, which is also consistent with our naive expectation that additional unit spin can lead to larger mass.

\begin{figure}
 \centering
 \includegraphics[totalheight=5cm,width=6cm]{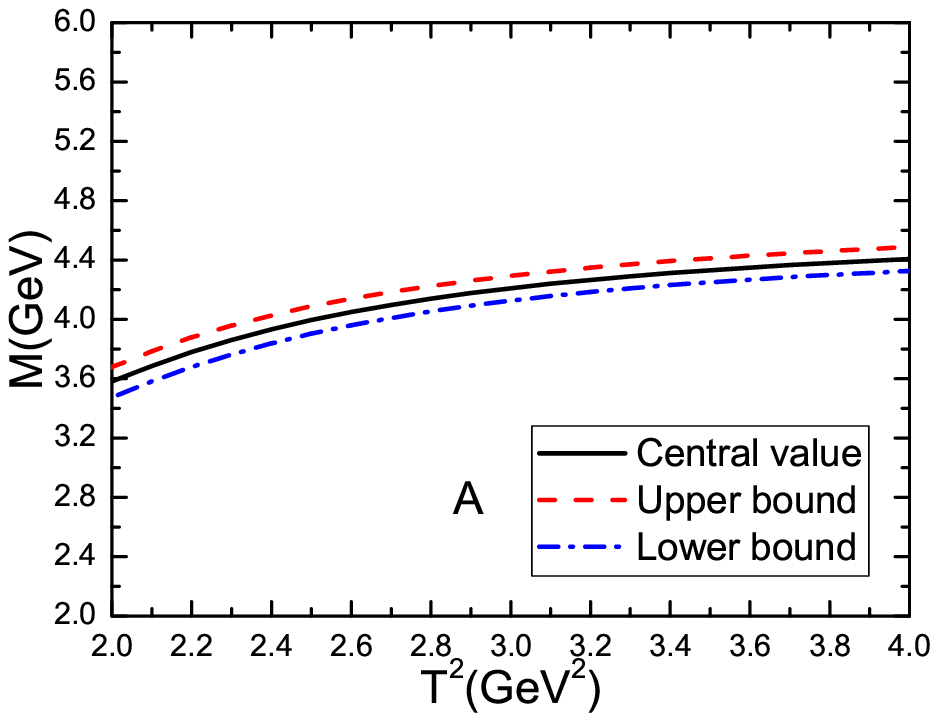}
 \includegraphics[totalheight=5cm,width=6cm]{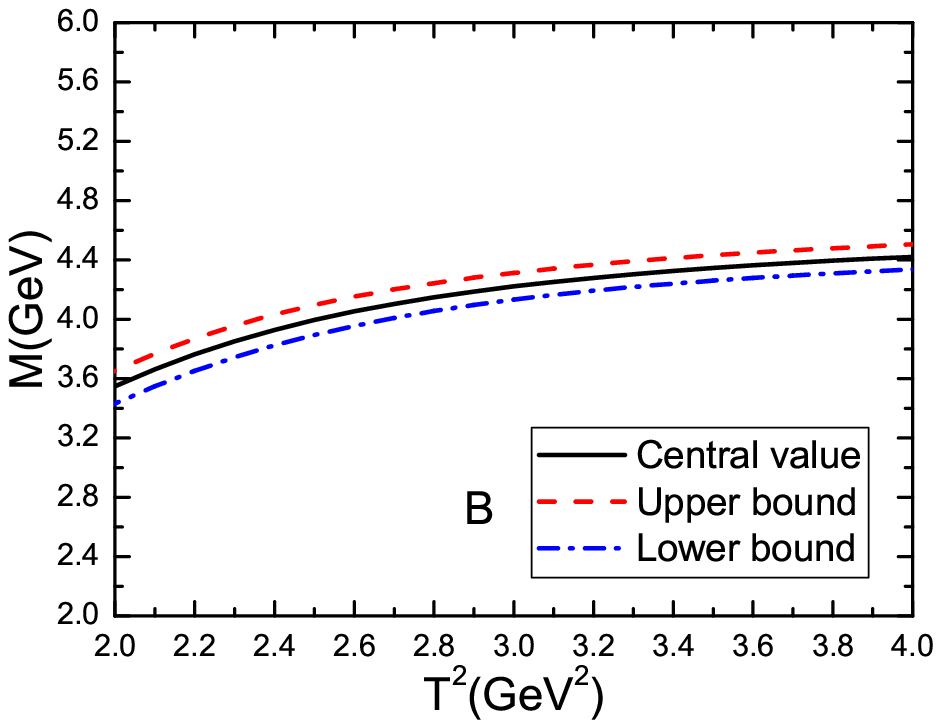}
 \includegraphics[totalheight=5cm,width=6cm]{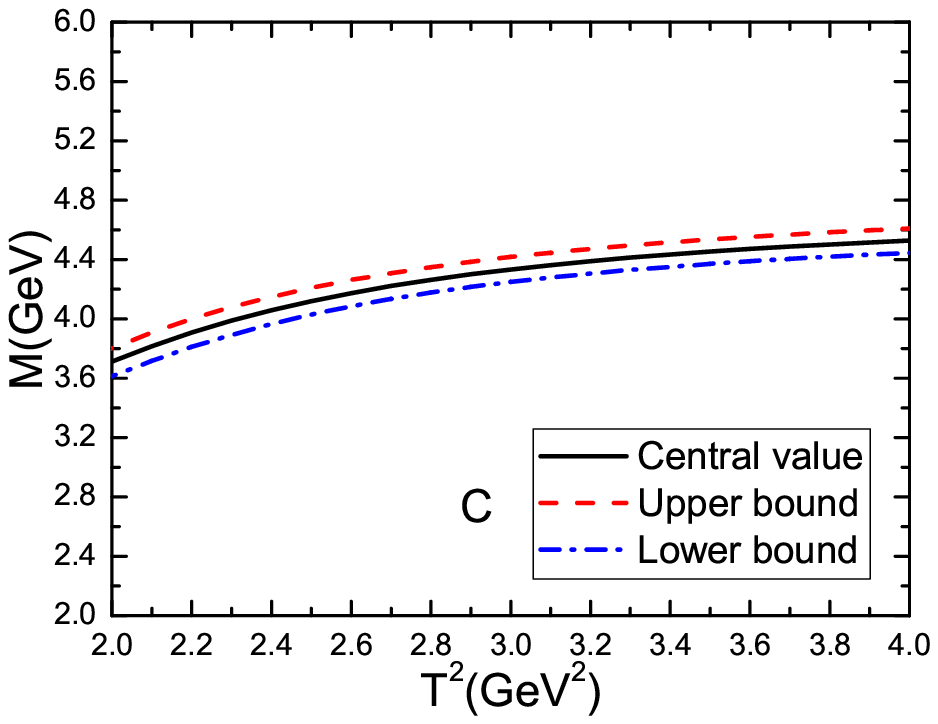}
 \includegraphics[totalheight=5cm,width=6cm]{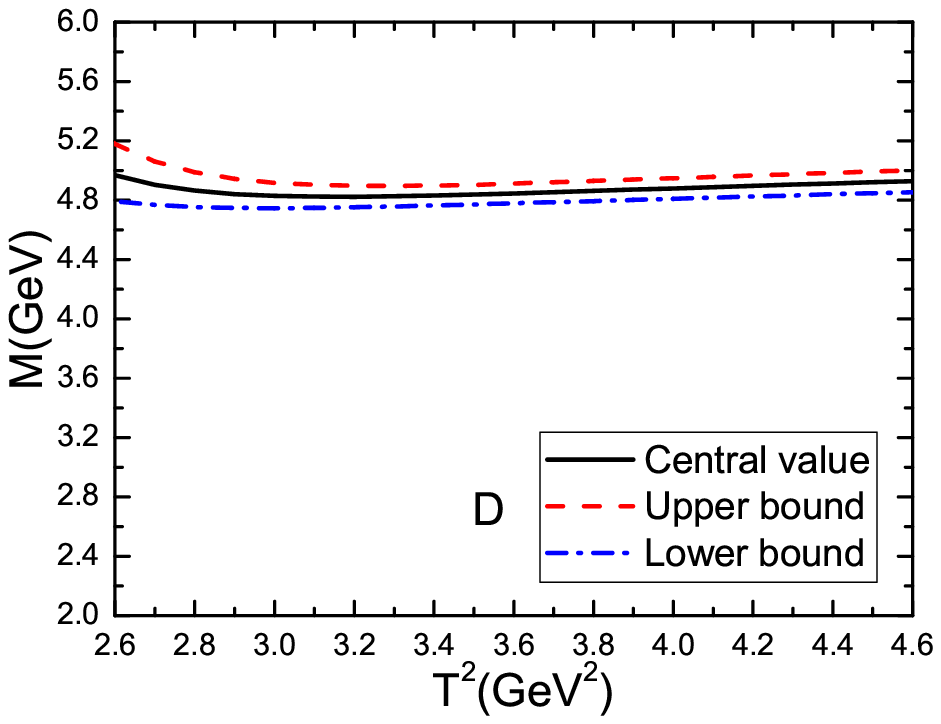}
  \caption{ The masses  of the pentaquark states  with variations of the Borel parameters $T^2$, where the $A$, $B$, $C$ and $D$ denote the pentaquark states
          $P_{00,-}$,   $P_{01,-}$,  $P_{00,+}$ and  $P_{01,+}$,  respectively.  }
\end{figure}

\begin{figure}
 \centering
 \includegraphics[totalheight=5cm,width=6cm]{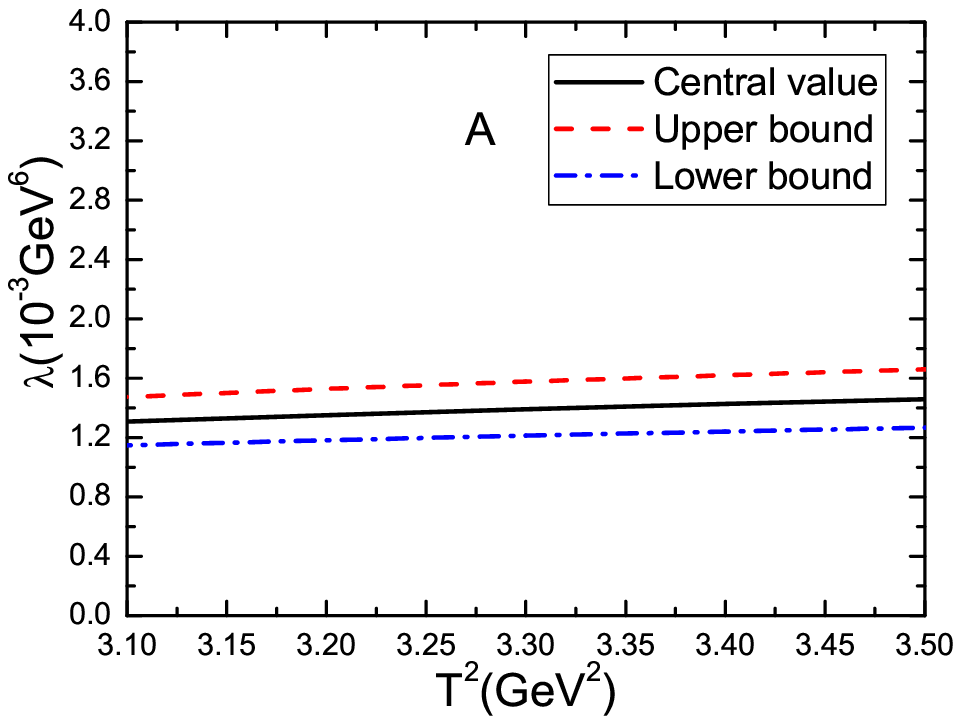}
 \includegraphics[totalheight=5cm,width=6cm]{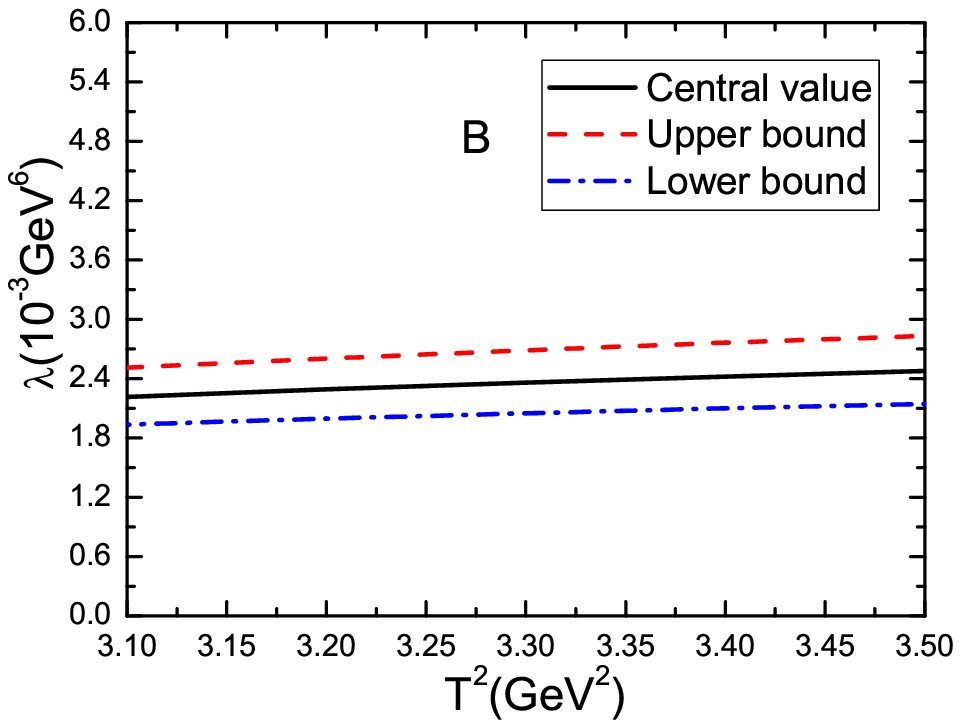}
 \includegraphics[totalheight=5cm,width=6cm]{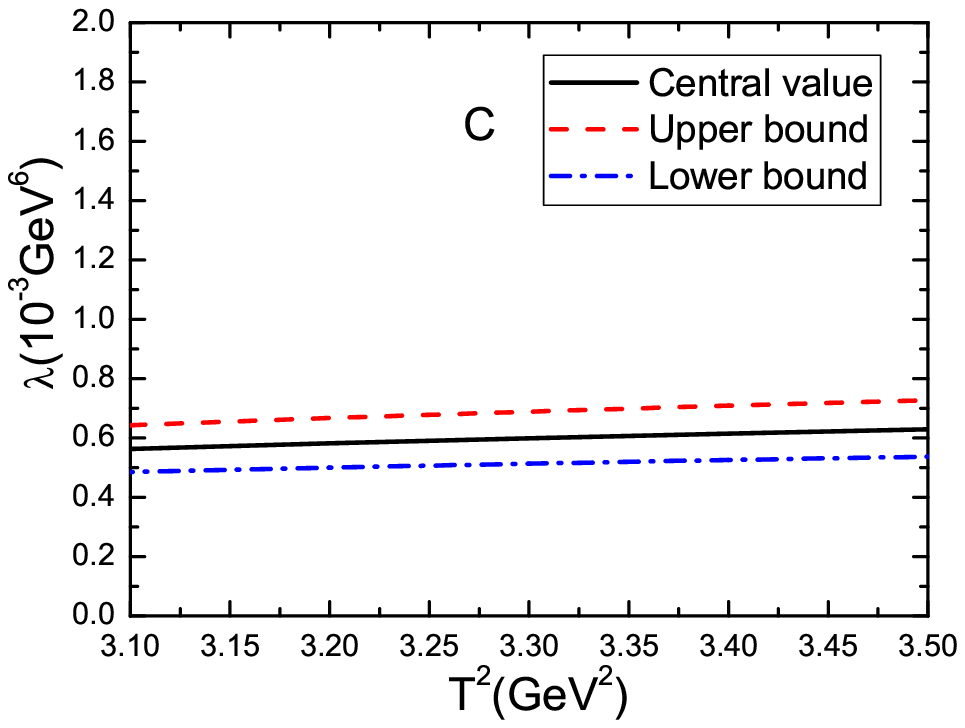}
 \includegraphics[totalheight=5cm,width=6cm]{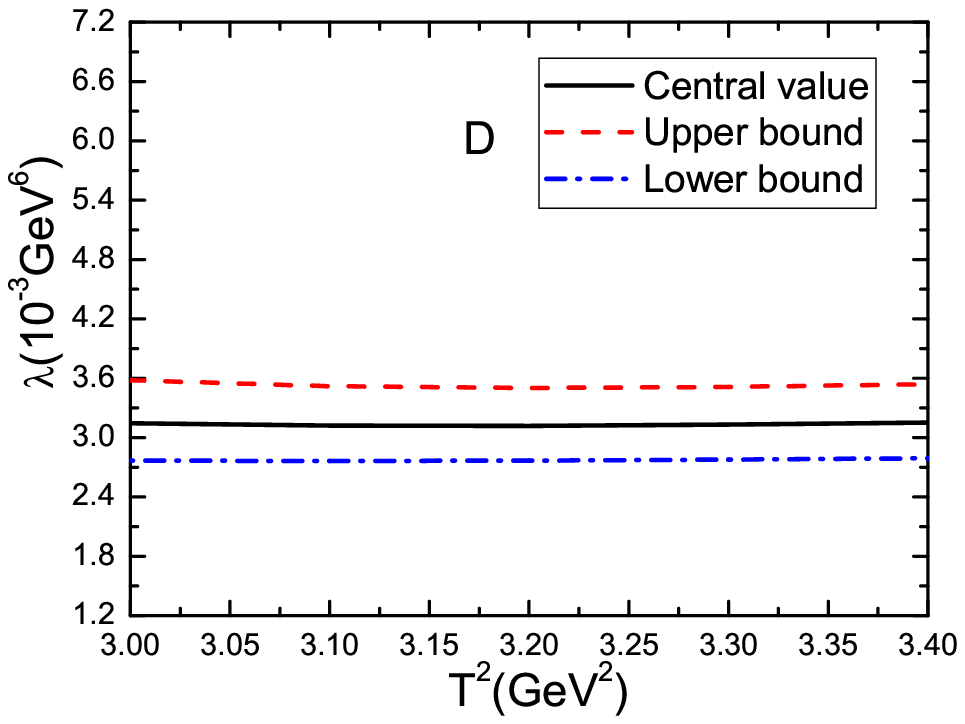}
 \caption{ The pole residues   of the pentaquark states  with variations of the Borel parameters $T^2$, where the $A$, $B$, $C$ and $D$ denote the pentaquark states
      $P_{00,-}$,  $P_{01,-}$,  $P_{00,+}$ and $P_{01,+}$, respectively.  }
\end{figure}

In the conventional QCD sum rules for the mesons, we usually take the continuum threshold parameters $\sqrt{s_0}=M_{\rm{gr}}+ (0.4-0.6)\,\rm{GeV}$  based on the assumption that the energy gap between the ground states and the first radial excited states is about $0.5\,\rm{GeV}$, where the gr denotes the ground states.
In Refs.\cite{WangHbaryon,WangLambda},  we  separate the contributions of the negative parity baryon states from that of the positive parity baryon states unambiguously,  study the  $J^P={1\over 2}^{\pm}$ and ${3\over 2}^{\pm}$ heavy, doubly-heavy and triply-heavy baryon states $qq^{\prime}Q$, $qQQ^\prime$, $QQ^{\prime}Q^{\prime\prime}$ with the QCD sum rules in a systematic way, the continuum threshold parameters $\sqrt{s_0}=M_{\rm{gr}}+ (0.6-0.8)\,\rm{GeV}$ work well, the experimental values of the masses can be well reproduced.

The pentaquark states are special baryon states, as they have one unit  baryon number.
In Ref.\cite{Wang1508},  we take the continuum threshold parameters
$\sqrt{s_0}= M_{P_c(4380/4450)}+(0.6-0.8)\,\rm{GeV}$, which also work well.
In this article, the optimal continuum threshold parameters are
$\sqrt{s_0}= M_{P}+(0.6-0.8)\,\rm{GeV}$. One may worry  that there maybe exist some contaminations from the high resonances and continuum states, as the spectroscopy of the pentaquark states is unclear in the present time. We should not be so pessimistic as  the high resonances and continuum states are greatly suppressed by the factor $\exp\left(-\frac{s}{T^2}\right)$. If we take the largest threshold parameters $s^0_{max}$ and the central  values of other
parameters, then
\begin{eqnarray}
\frac{\exp\left(-\frac{s^0_{max}}{T^2}\right)}{\exp\left(-\frac{M_P^2}{T^2}\right)} &\leq & 10\%\, ,
\end{eqnarray}
the contaminations are greatly suppressed compared to the ground states, so the predictive ability cannot be impaired remarkably.  The present predictions can be confronted with the experimental data in the future.

In Fig.4, we plot the contributions of the pole terms   with
variations of the continuum threshold parameters $\sqrt{s_0}$ and Borel parameters $T^2$  for the pentaquark states  $P_{00,-}$, $P_{01,-}$,
    $P_{00,+}$ and   $P_{01,+}$ at the energy scales presented in Table 2.
From the figure, we can see that the   pole  contributions decrease quickly and monotonously with the increase of the Borel parameters  for the
pentaquark states $P_{00,-}$, $P_{01,-}$ and   $P_{00,+}$,
 the pole contributions reach $50\%$ at  $T^2\approx 3.3\,\rm{GeV}^2$ with the central values of the
continuum threshold parameters. For the pentaquark state $P_{01,+}$, the integral
\begin{eqnarray}
\int_{4m_c^2}^{s_0}ds \left[\sqrt{s}\rho_{01,QCD}^1(s)-m_c\widetilde{\rho}_{01,QCD}^0(s)\right]\exp\left( -\frac{s}{T^2}\right) &<& 0 \, ,
\end{eqnarray}
at the value $T^2< 2.0\,\rm{GeV}^2$, which magnifies itself by the strange behavior of the pole contribution in Fig.4-D; while at the value
$T^2> 2.3\,\rm{GeV}^2$,  the integral is positive, the   pole  contribution  decreases quickly and monotonously with the increase of the Borel parameter, and reaches $50\%$ at the $T^2\approx 3.6\,\rm{GeV}^2$.   We can draw the conclusion tentatively that the  convergent behavior of the $P_{01,+}$ differs from that of the
$P_{00,-}$, $P_{01,-}$ and   $P_{00,+}$ significantly, as it has much larger pole contribution in the Borel window, see Table 1. On the other hand, if we try to obtain smaller pole contribution, say about $(40-60)\%$ by choosing larger Borel parameters, the energy scale formula in Eq.(31) cannot be satisfied.    From Fig.2-D, we can see that the  Borel platform of the predicted mass $M_{01,+}$ appears as the minimum value, and the platform is very flat, which originates from the special convergent behavior in the operator product expansion. The negative integral  at the value $T^2< 2.0\,\rm{GeV}^2$ or  $\sqrt{T^2}< 1.4\,\rm{GeV}$ shown in Eq.(39)  is acceptable, as  the optimal  energy scale $ \mu =3.2\,\rm{GeV}\gg 1.4\,\rm{GeV}$ (see Table 2 or Fig.5-D), the value $T^2< 2.0\,\rm{GeV}^2$ or  $\sqrt{T^2}< 1.4\,\rm{GeV}$ is out of the allowed region of the Borel parameter $T^2=(3.0-3.4)\,\rm{GeV}^2$, where the four criteria of the QCD sum rules can be satisfied. If we take into account the higher excited states besides the ground state, a larger continuum threshold $s_0$ is needed, therefore larger Borel parameter $T^2$ is needed to magnify  the contributions of the higher excited states, then integral in Eq.(39) is also positive. So in the allowed region of the Borel parameter, the integral in Eq.(39) is positive.
The continuum contributions can be approximated as
\begin{eqnarray}
 \int_{s_0}^{\infty}ds \rho_{H}(s) \exp\left( -\frac{s}{T^2}\right)&=&\int_{s_0}^{\infty}ds \left[\sqrt{s}\rho_{01,QCD}^1(s)-m_c\widetilde{\rho}_{01,QCD}^0(s)\right]\exp\left( -\frac{s}{T^2}\right)   \, ,
\end{eqnarray}
where the $\rho_{H}(s)$ denotes the hadronic spectral density. At the value $T^2< 2.0\,\rm{GeV}^2$, the continuum contributions are greatly depressed, for example, $\exp\left( -\frac{s_0}{T^2}\right)\leq \exp\left( -\frac{5.4^2}{2}\right)=4.7\times 10^{-7} $, and it is   out of the  allowed region of the Borel parameter.
Furthermore, in the limit $T^2\rightarrow \infty$ or in the local limit,
\begin{eqnarray}
\int_{4m_c^2}^{s_0}ds \left[\sqrt{s}\rho_{01,QCD}^1(s)-m_c\widetilde{\rho}_{01,QCD}^0(s)\right]\exp\left( -\frac{s}{T^2}\right) &\rightarrow& \int_{4m_c^2}^{s_0}ds \left[\sqrt{s}\rho_{01,QCD}^1(s)-m_c\widetilde{\rho}_{01,QCD}^0(s)\right]  \nonumber\\
&>& 0\, ,
\end{eqnarray}
a positive spectral density can be warranted.

\begin{figure}
 \centering
 \includegraphics[totalheight=5cm,width=6cm]{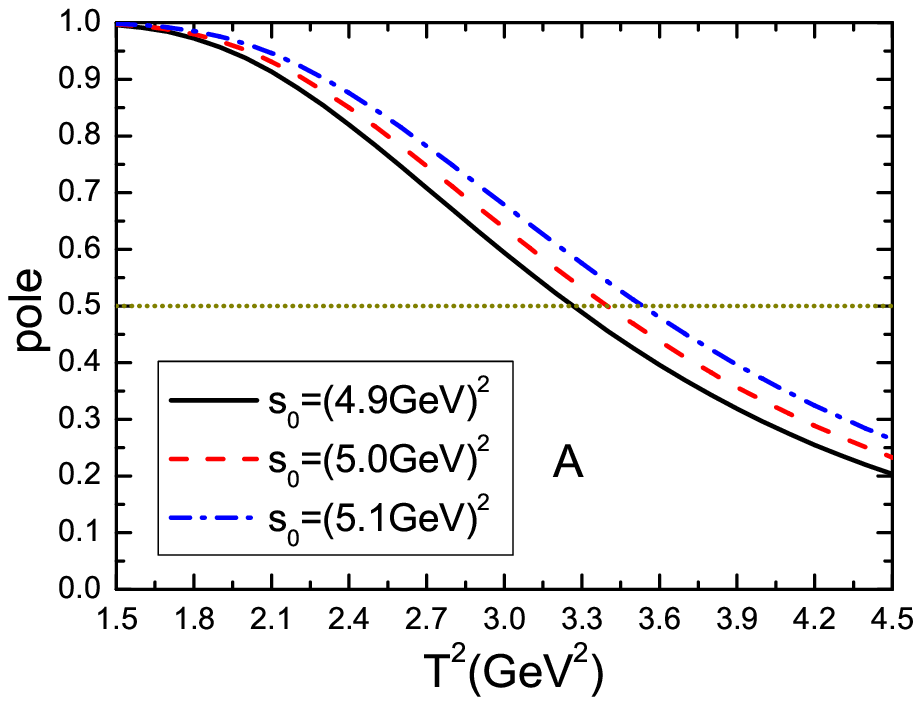}
 \includegraphics[totalheight=5cm,width=6cm]{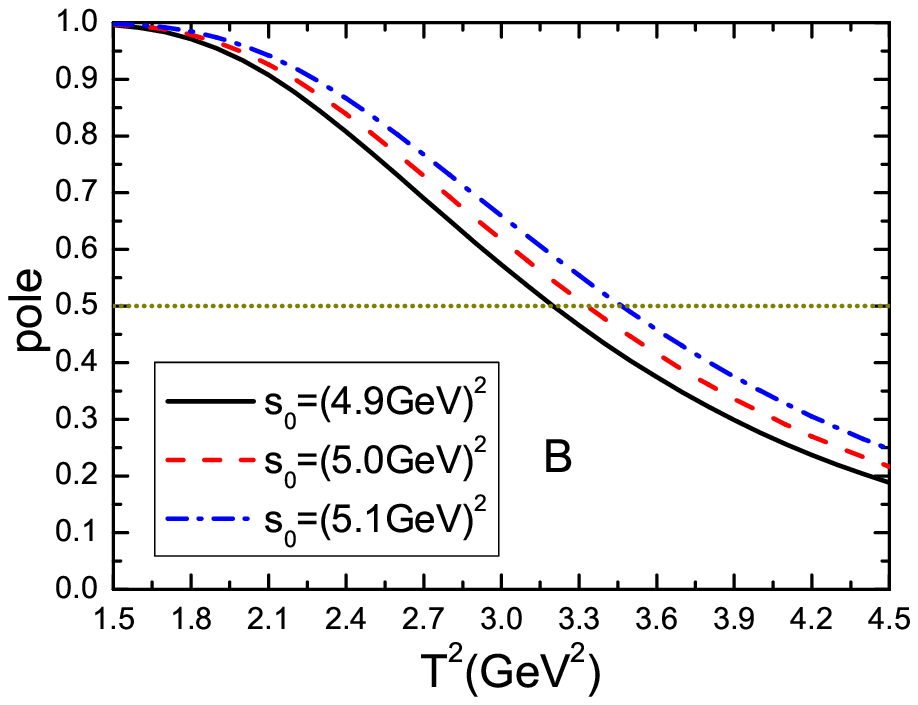}
 \includegraphics[totalheight=5cm,width=6cm]{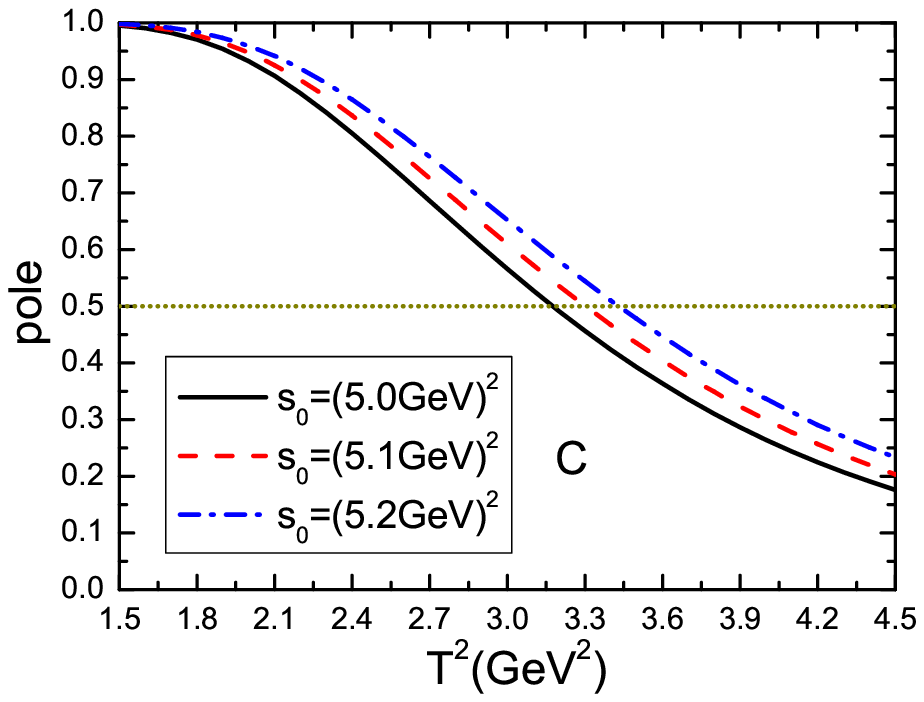}
 \includegraphics[totalheight=5cm,width=6cm]{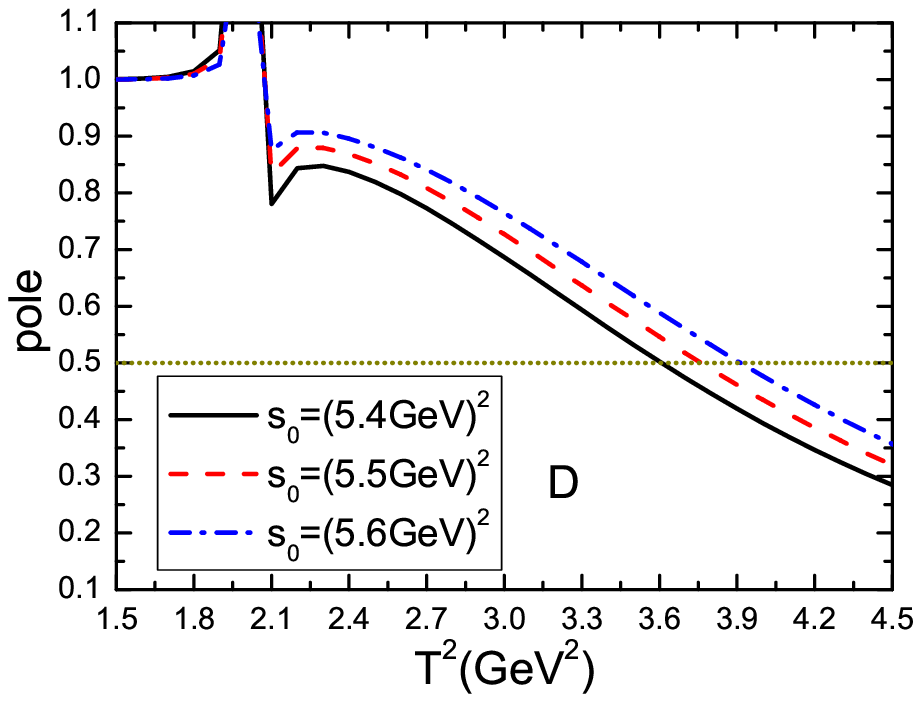}
  \caption{ The pole contributions  of the pentaquark states  with variations of the Borel parameters $T^2$, where the $A$, $B$, $C$ and $D$ denote the pentaquark states
          $P_{00,-}$,   $P_{01,-}$,  $P_{00,+}$ and  $P_{01,+}$,  respectively.  }
\end{figure}

\begin{table}
\begin{center}
\begin{tabular}{|c|c|c|c|c|c|c|c|}\hline\hline
           &$T^2 (\rm{GeV}^2)$ &$\sqrt{s_0} (\rm{GeV})$ &pole         &$\langle\bar{q}q\rangle^3$ &$\langle\bar{q}g_s\sigma Gq\rangle^2$ \\ \hline
$P_{00,-}$ &$3.1-3.5$          &$5.0\pm0.1$             &$(43-64)\%$  &$(12-17)\%$                &$(2-3)\%$       \\ \hline
$P_{01,-}$ &$3.1-3.5$          &$5.0\pm0.1$             &$(40-62)\%$  &$(13-18)\%$                &$(3-5)\%$        \\ \hline
$P_{00,+}$ &$3.1-3.5$          &$5.1\pm0.1$             &$(39-62)\%$  &$(14-20)\%$                &$(3-4)\%$    \\ \hline
$P_{01,+}$ &$3.0-3.4$          &$5.5\pm0.1$             &$(56-76)\%$  &$-(6-12)\%$                &$(3-6)\%$        \\ \hline
 \hline
\end{tabular}
\end{center}
\caption{ The Borel parameters, continuum threshold parameters, pole contributions, contributions of the vacuum condensates of dimension 9 ($\langle\bar{q}q\rangle^3$) and dimension 10 ($\langle\bar{q}g_s\sigma Gq\rangle^2$). }
\end{table}

\begin{table}
\begin{center}
\begin{tabular}{|c|c|c|c|c|c|c|c|}\hline\hline
            & $T^2 (\rm{GeV}^2)$ & $\sqrt{s_0} (\rm{GeV})$  & $\mu (\rm{GeV})$  & $M_{P}(\rm{GeV})$   & $\lambda_{P}(\rm{GeV}^6)$ \\ \hline
$P_{00,-}$  & $3.1-3.5$          & $5.0\pm0.1$              & $2.3$             & $4.29\pm0.13$       & $(1.39\pm0.26)\times10^{-3}$ \\ \hline
$P_{01,-}$  & $3.1-3.5$          & $5.0\pm0.1$              & $2.4$             & $4.30\pm0.13$       & $(2.36\pm0.45)\times10^{-3}$\\ \hline
$P_{00,+}$  & $3.1-3.5$          & $5.1\pm0.1$              & $2.5$             & $4.41\pm0.13$       & $(0.60\pm0.12)\times10^{-3}$\\ \hline
$P_{01,+}$  & $3.0-3.4$          & $5.5\pm0.1$              & $3.2$             & $4.82\pm0.08$       & $(3.11\pm0.37)\times10^{-3}$\\ \hline
 \hline
\end{tabular}
\end{center}
\caption{ The Borel parameters, continuum threshold parameters,  energy scales,  masses and pole residues of the pentaquark states.  }
\end{table}

\begin{figure}
 \centering
 \includegraphics[totalheight=5cm,width=6cm]{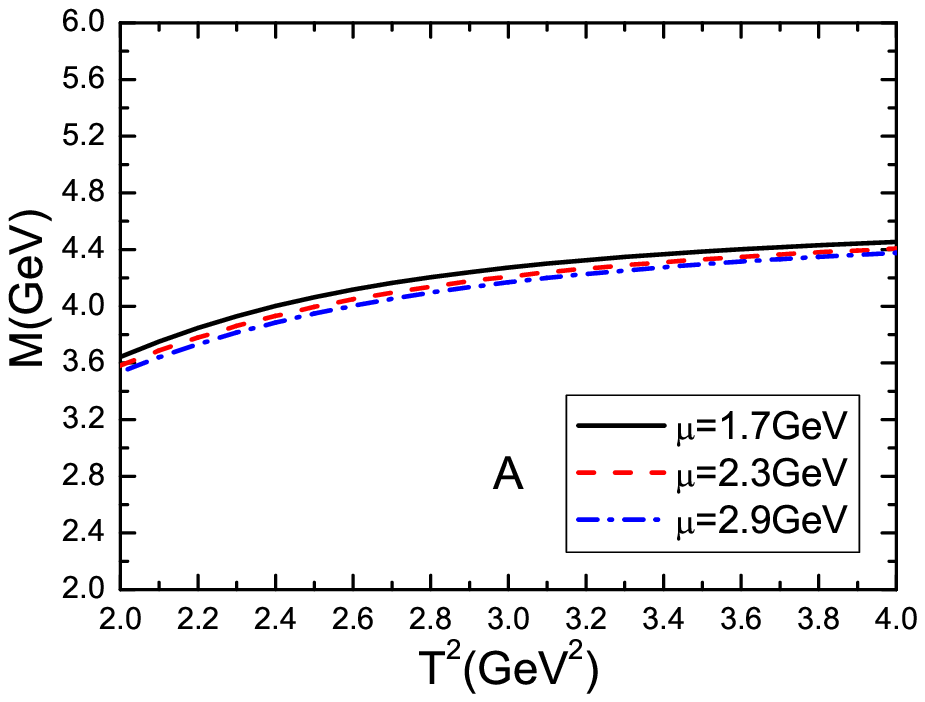}
 \includegraphics[totalheight=5cm,width=6cm]{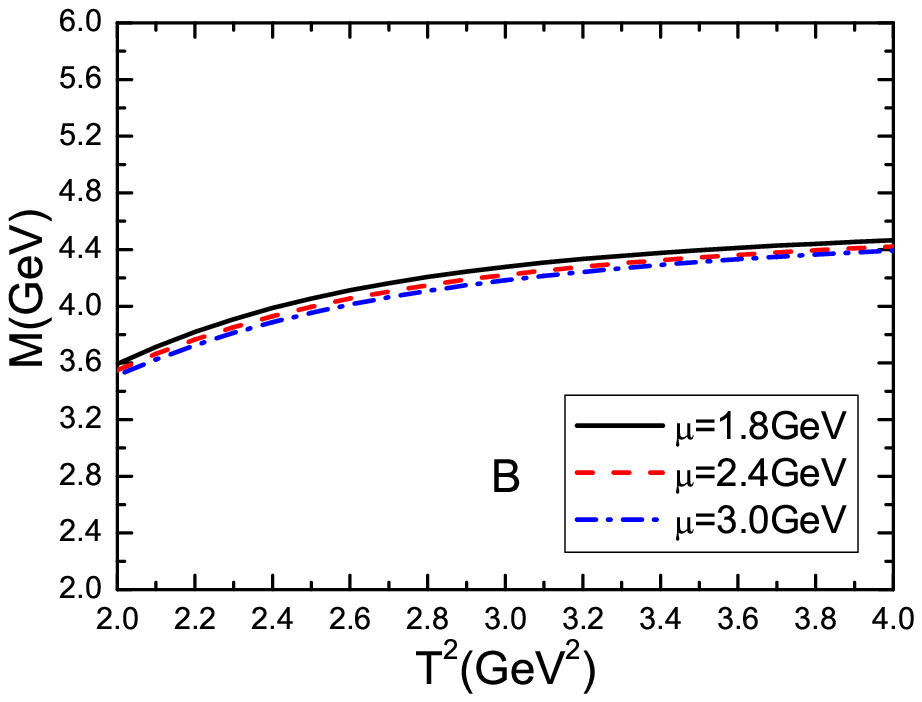}
 \includegraphics[totalheight=5cm,width=6cm]{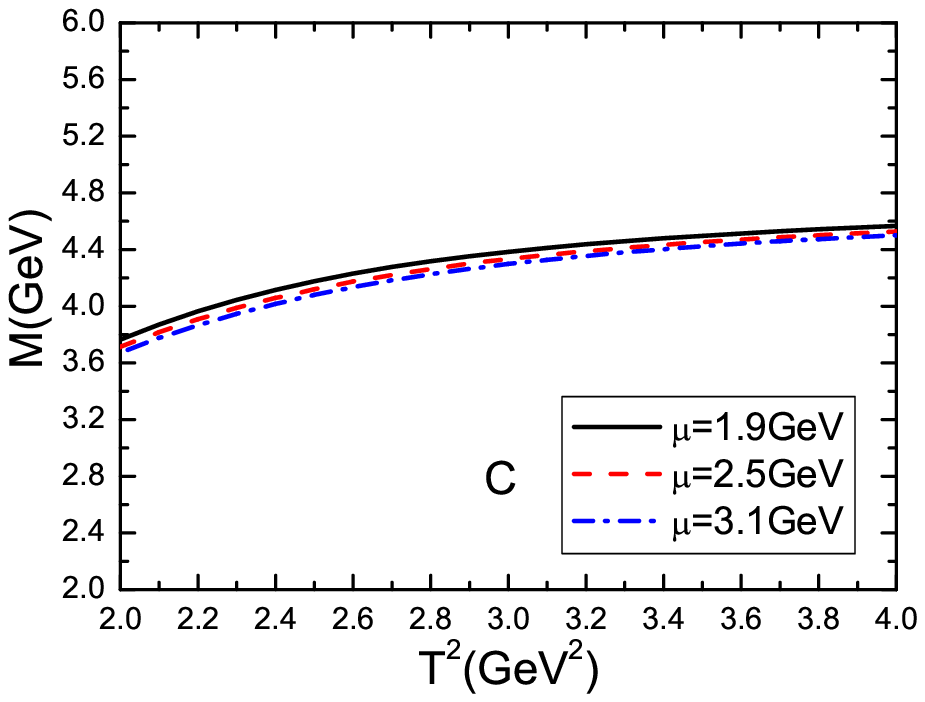}
 \includegraphics[totalheight=5cm,width=6cm]{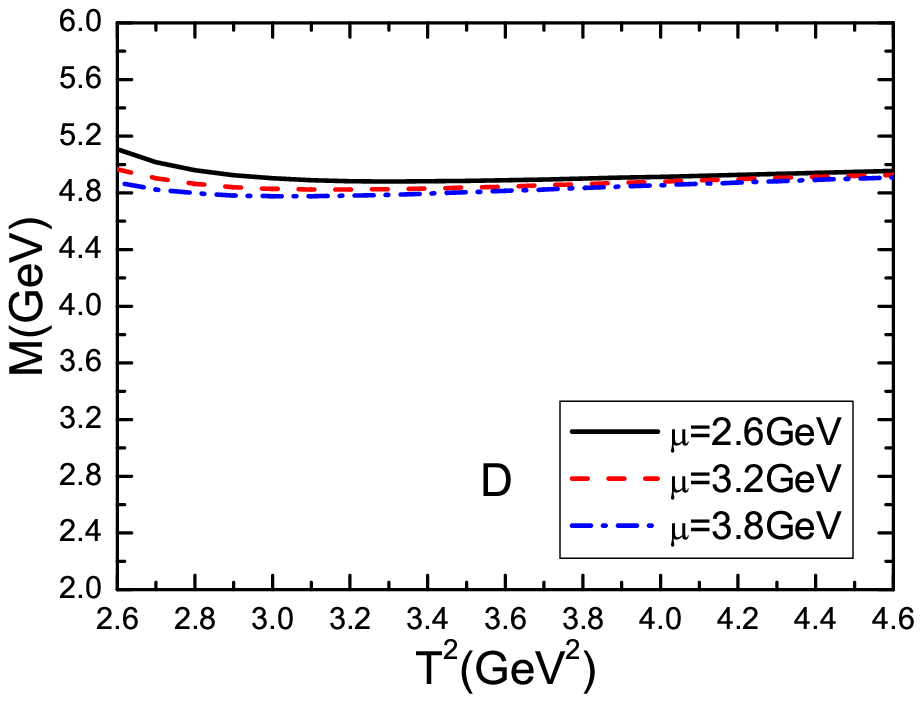}
  \caption{ The masses  of the pentaquark states  with variations of the Borel parameters $T^2$ and energy scales $\mu$, where the $A$, $B$, $C$ and $D$ denote the pentaquark states
          $P_{00,-}$,   $P_{01,-}$,  $P_{00,+}$ and  $P_{01,+}$,  respectively.  }
\end{figure}

In this article, the contributions $ D_{j_L j_H,i,\pm}$ of the   vacuum condensates of dimension-$i$ with $i=0,\,3,\,4,\,5,\,6,\,8,\,9,\,10$ are defined by
\begin{eqnarray}
 D_{j_L j_H,i,+} &=& \frac{\int_{4m_c^2}^{s_0}ds \left[\sqrt{s}\rho_{j_L j_H,i}^1(s)-m_c\widetilde{\rho}_{j_L j_H,i}^0(s)\right]\,\exp\left( -\frac{s}{T^2}\right)}{\int_{4m_c^2}^{s_0}ds \left[\sqrt{s}\rho_{j_L j_H,QCD}^1(s)-m_c\widetilde{\rho}_{j_L j_H,QCD}^0(s)\right]\exp\left( -\frac{s}{T^2}\right)}\, ,\\
D_{j_L j_H,i,-} &=&\frac{\int_{4m_c^2}^{s_0}ds \left[\sqrt{s}\rho_{j_L j_H,i}^1(s)+m_c\widetilde{\rho}_{j_L j_H,i}^0(s)\right]\,\exp\left( -\frac{s}{T^2}\right)}{\int_{4m_c^2}^{s_0}ds \left[\sqrt{s}\rho_{j_L j_H,QCD}^1(s)+m_c\widetilde{\rho}_{j_L j_H,QCD}^0(s)\right]\exp\left( -\frac{s}{T^2}\right)}\, ,
\end{eqnarray}
which do not warrant the contributions $ D_{0 0,i,-}$, $ D_{0 1,i,-}$, $ D_{0 0,i,+}$ and $ D_{0 1,i,+}$ have the same positive (or negative) sign, see Table 1.
On the other hand, if we define  the contributions $ \overline{D}_{j_L j_H,i,\pm}$ of the vacuum condensates of  dimension-$i$ by
\begin{eqnarray}
 \overline{D}_{j_L j_H,i,+} &=& \frac{\int_{4m_c^2}^{\infty}ds \left[\sqrt{s}\rho_{j_L j_H,i}^1(s)-m_c\widetilde{\rho}_{j_L j_H,i}^0(s)\right]\,\exp\left( -\frac{s}{T^2}\right)}{\int_{4m_c^2}^{\infty}ds \left[\sqrt{s}\rho_{j_L j_H,QCD}^1(s)-m_c\widetilde{\rho}_{j_L j_H,QCD}^0(s)\right]\exp\left( -\frac{s}{T^2}\right)}\, ,\\
\overline{D}_{j_L j_H,i,-} &=&\frac{\int_{4m_c^2}^{\infty}ds \left[\sqrt{s}\rho_{j_L j_H,i}^1(s)+m_c\widetilde{\rho}_{j_L j_H,i}^0(s)\right]\,\exp\left( -\frac{s}{T^2}\right)}{\int_{4m_c^2}^{\infty}ds \left[\sqrt{s}\rho_{j_L j_H,QCD}^1(s)+m_c\widetilde{\rho}_{j_L j_H,QCD}^0(s)\right]\exp\left( -\frac{s}{T^2}\right)}\, ,
\end{eqnarray}
contributions of the terms $\sqrt{s}\rho_{j_L j_H,i}^1(s)$  are greatly enhanced compared to the terms $m_c\widetilde{\rho}_{j_L j_H,i}^0(s)$, which maybe lead to the contributions $ \overline{D}_{0 0,i,-}$, $ \overline{D}_{0 1,i,-}$, $\overline{ D}_{0 0,i,+}$ and $ \overline{D}_{0 1,i,+}$ have the same positive (or negative) sign. However, we have to take into account the contributions of the high resonances and continuum states at the phenomenological side in the QCD sum rules.

The correlation functions $\Pi_{j_Lj_H}(p)$ can be written as
\begin{eqnarray}
\Pi_{j_Lj_H}(p)&=&\sum_n C_n(p^2,\mu)\langle{\mathcal{O}}_n(\mu)\rangle=\int_{4m^2_c(\mu)}^\infty ds \frac{\rho_{QCD}(s,\mu)}{s-p^2} \nonumber\\
&=&\int_{4m^2_c(\mu)}^{s_0} ds \frac{\rho_{QCD}(s,\mu)}{s-p^2}+\int_{s_0}^\infty ds \frac{\rho_{QCD}(s,\mu)}{s-p^2} \, ,
\end{eqnarray}
at the QCD side, where the $C_n(p^2,\mu)$ are the Wilson coefficients and the $\langle{\mathcal{O}}_n(\mu)\rangle$ are the vacuum condensates of dimension-$n$.
 At the energy scale  $\mu\gg \Lambda_{QCD}$, the short-distance contributions at  $p^2>\mu^2$ are included in the coefficients
$C_n(p^2,\mu)$, the long-distance contributions at $p^2<\mu^2$ are absorbed into the vacuum condensates  $\langle{\mathcal{O}}_n(\mu)\rangle$.
The correlation functions $\Pi_{j_Lj_H}(p)$ are scale independent,
\begin{eqnarray}
\frac{d}{d\mu}\Pi_{j_Lj_H}(p)&=&0\, ,
\end{eqnarray}
which does not warrant
\begin{eqnarray}
\frac{d}{d\mu}\int_{4m^2_c(\mu)}^{s_0} ds \frac{\rho_{QCD}(s,\mu)}{s-p^2}\rightarrow 0 \, ,
\end{eqnarray}
 due to the following two reasons in the present QCD sum rules:\\
$\bf 1.$ Perturbative corrections are not available, the higher dimensional vacuum condensates are factorized into lower dimensional ones therefore  the energy scale dependence of the higher dimensional vacuum condensates is modified;\\
$\bf 2.$ Truncations $s_0$ set in, the correlation between the threshold $4m^2_c(\mu)$ and continuum threshold $s_0$ is unknown,  the quark-hadron duality is an assumption.

We cannot obtain energy scale independent QCD sum rules even if perturbative corrections are available, for example, in the case of the conventional heavy-light mesons \cite{Wangdecaycon}, but we have typical energy scales which  characterize the five-quark systems  $uudc\bar{c}$ according to Eqs.(28-31) and serve as the optimal energy scales of the QCD spectral densities.
In Fig.5, we plot the predicted masses   with
variations of the  Borel parameters $T^2$  and energy scales $\mu$ for the pentaquark states  $P_{00,-}$, $P_{01,-}$,
    $P_{00,+}$ and   $P_{01,+}$, respectively.  From the figure, we can see that the predicted masses decrease monotonously with increase of the energy scales $\mu$.
If we take the central values of the Borel parameters presented in Table 2, the uncertainties induced by the uncertainties $\delta\mu=\pm0.6\,\rm{GeV}$ are about ${}_{-0.04}^{+0.06}\,\rm{GeV}$, ${}_{-0.04}^{+0.05}\,\rm{GeV}$, ${}_{-0.03}^{+0.05}\,\rm{GeV}$ and  ${}_{-0.04}^{+0.06}\,\rm{GeV}$ for the
pentaquark states $P_{00,-}$, $P_{01,-}$,  $P_{00,+}$ and $P_{01,+}$, respectively.   We can draw the conclusion tentatively that the  uncertainties induced by the uncertainties of the  energy scales in the vicinity of the optimal values are small. In calculations, we search for  the optimal  Borel parameters $T^2$ and continuum threshold parameters $s_0$  to reproduce the masses of the pentaquark states to
satisfy the energy scale formula in Eq.(31). In other words, we take the  energy scale formula in Eq.(31) as a constraint, and do not take the energy scales of the QCD spectral densities as input parameters.

The diquark-diquark-antiquark type current with special quantum numbers couples potentially  to  special pentaquark states. The current can be re-arranged both in the color and Dirac-spinor  spaces, and changed  to a current as a special superposition of  the color singlet  baryon-meson type currents.   The baryon-meson type currents couple potentially  to the baryon-meson pairs. The
diquark-diquark-antiquark type pentaquark state can be taken as a special superposition of a series of  baryon-meson pairs, and embodies  the net effects. The decays to its components (baryon-meson pairs) are Okubo-Zweig-Iizuka super-allowed, but the re-arrangements in the color-space are non-trivial \cite{Nielsen3900}.

In the following, we perform Fierz re-arrangement  to the currents $J_{00}$ and $J_{01}$ both in the color and Dirac-spinor  spaces to   obtain the results,
\begin{eqnarray}
J_{00} &=&\frac{1}{4}\mathcal{S}u\,\bar{c} c-\frac{1}{4}\mathcal{S}c\,\bar{c} u+\frac{1}{4}\mathcal{S}\gamma^\alpha u\,\bar{c}\gamma_\alpha c-\frac{1}{4}\mathcal{S}\gamma^\alpha c\,\bar{c}\gamma_\alpha u- \frac{1}{8}\mathcal{S}\sigma_{\alpha\beta} u\,\bar{c} \sigma^{\alpha\beta}c+\frac{1}{8}\mathcal{S}\sigma_{\alpha\beta} c\,\bar{c}\sigma^{\alpha\beta} u\nonumber\\
&& +\frac{1}{4}\mathcal{S}\gamma^\alpha \gamma_5 u\,\bar{c}\gamma_\alpha\gamma_5 c-\frac{1}{4}\mathcal{S}\gamma^\alpha\gamma_5 c\,\bar{c}\gamma_\alpha\gamma_5 u
-\frac{i}{4}\mathcal{S} \gamma_5 u\,\bar{c}i\gamma_5 c+\frac{i}{4}\mathcal{S}\gamma_5 c\,\bar{c}i\gamma_5 u\, ,
\end{eqnarray}
\begin{eqnarray}
J_{01} &=&-\mathcal{S}u\,\bar{c} c-\mathcal{S}c\,\bar{c} u-\frac{1}{2}\mathcal{S}\gamma^\alpha u\,\bar{c}\gamma_\alpha c-\frac{1}{2}\mathcal{S}\gamma^\alpha c\,\bar{c}\gamma_\alpha u +\frac{1}{2}\mathcal{S}\gamma^\alpha \gamma_5 u\,\bar{c}\gamma_\alpha\gamma_5 c+\frac{1}{2}\mathcal{S}\gamma^\alpha\gamma_5 c\,\bar{c}\gamma_\alpha\gamma_5 u \nonumber\\
&& -i\mathcal{S} \gamma_5 u\,\bar{c}i\gamma_5 c-i\mathcal{S}\gamma_5 c\,\bar{c}i\gamma_5 u\, ,
\end{eqnarray}
  where we  use   the notations  $\mathcal{S}\Gamma c=\varepsilon^{ijk}u^T_i C \gamma_5 d_j \Gamma c_k$ and $\mathcal{S}\Gamma u=\varepsilon^{ijk}u^T_i C \gamma_5 d_j \Gamma u_k$ for simplicity, here the $\Gamma$ denotes the Dirac matrixes.

  The components $\mathcal{S}(x)\Gamma c(x) \bar{c}(x)\Gamma^{\prime}u(x)$ and $\mathcal{S}(x)\Gamma u(x) \bar{c}(x)\Gamma^{\prime}c(x)$ couple potentially to the baryon-meson pairs. The revelent thresholds are $M_{\eta_c p}=3.922\,\rm{GeV}$, $M_{J/\psi p}=4.035\,\rm{GeV}$, $M_{\Lambda_c^+ \bar{D}^0}=4.151\,\rm{GeV}$,
   $M_{\Lambda_c^+ \bar{D}^{*0}}=4.293\,\rm{GeV}$,
  $M_{\chi_{c0} p}=4.353\,\rm{GeV}$,  $M_{\eta_c N(1440)}=4.414\,\rm{GeV}$,  $M_{\chi_{c1} p}=4.449\,\rm{GeV}$, $M_{\Lambda_c^+(2595) \bar{D}^0}=4.457\,\rm{GeV}$,
  $M_{h_{c} p}=4.463\,\rm{GeV}$, $M_{\Lambda_c^+(2595) \bar{D}^{*0}}=4.599\,\rm{GeV}$,
  $M_{\Lambda_c^+ \bar{D}^0_0(2400)}=4.604\,\rm{GeV}$,
  $M_{\Lambda_c^+ \bar{D}^0_1(2420)}=4.708\,\rm{GeV}$, $M_{\Lambda_c^+ \bar{D}^0_1(2430)}=4.713\,\rm{GeV}$
    \cite{PDG}. After taking into account the currents-hadrons duality, we obtain the Okubo-Zweig-Iizuka super-allowed decays,
\begin{eqnarray}
P_{00,-}(4290) &\to& p J/\psi  \, , \,  p\eta_c \, , \,\Lambda_c^+ \bar{D}^{0} \, , \\
P_{00,+}(4410) &\to& p J/\psi  \, , \, \Lambda_c^+ \bar{D}^{*0}\, , \, p\eta_c \, , \,\Lambda_c^+ \bar{D}^{0}\, , \, p\chi_{c0} \, , \\
P_{01,-}(4300) &\to& p J/\psi  \, , \, \Lambda_c^+ \bar{D}^{*0}\, , \, p\eta_c \, , \,\Lambda_c^+ \bar{D}^{0} \, , \\
P_{01,+}(4820) &\to& p J/\psi  \, , \, \Lambda_c^+ \bar{D}^{*0}\, , \,\Lambda_c^+(2595) \bar{D}^{*0}\, , \, p\eta_c \, , \, N(1440)\eta_c \, , \, p\chi_{c1}\, , \,\Lambda_c^+ \bar{D}^{0}\, , \,\Lambda_c^+(2595) \bar{D}^{0}\, , \,  \nonumber\\
&&\Lambda_c^+ \bar{D}^{0}_1(2420/2430)\, , \, p\chi_{c0} \, , \, \Lambda_c^+ \bar{D}^{0}_0(2400)\, ,
\end{eqnarray}
where we add the   masses of the pentaquark states in the brackets. We can search for the $P_{00,-}(4290)$, $P_{00,+}(4410)$, $P_{01,-}(4300)$ and $P_{01,+}(4820)$ in the those decays in the future.

\section{Summary and discussions}
In this article, we present the scalar-diquark-scalar-diquark-antiquark type and scalar-diquark-axialvector-diquark-antiquark type pentaquark configurations in the diquark model firstly, then construct  both the scalar-diquark-scalar-diquark-antiquark type and scalar-diquark-axialvector-diquark-antiquark type interpolating currents,  and  study the masses and pole residues of the  $J^P={\frac{1}{2}}^\pm$   hidden-charm pentaquark states   in details with the QCD sum rules by calculating the contributions of the vacuum condensates up to dimension-10 in the operator product expansion. In calculations,  we use the  formula $\mu=\sqrt{M^2_{P}-(2{\mathbb{M}}_c)^2}$  to determine  the energy scales of the QCD spectral densities.     We can search for the pentaquark states $P_{00,-}(4290)$,
$P_{00,+}(4410)$, $P_{01,-}(4300)$ and $P_{01,+}(4820)$ in the decays listed in Eqs.(51-54), and confront the present predictions of the masses
 to the experimental data in the future.

The  LHCb collaboration studied the $\Lambda_b^0\to J/\psi  p K^-$ decays and observed  two pentaquark candidates $P_c(4380)$ and $P_c(4450)$ in the $J/\psi p$ mass spectrum  \cite{LHCb-4380}.  The $\Lambda_b^0$ can be well interpolated by the current $J(x)=\varepsilon^{ijk}u^T_i(x) C\gamma_5 d_j(x) b_k(x)$ \cite{WangLambda}, the $u$ and $d$ quark in the $\Lambda_b^0$ form a scalar diquark $[ud]_{\bar{3}}$ in color antitriplet, the decays $\Lambda_b^0\to J/\psi  p K^-$ take  place through the mechanism $\Lambda_b^0([ud] b )\to [ud] c\bar{c}s\to [ud] c\bar{c}u\bar{u}s\to P_c^+([ud] [uc]\bar{c})K^-(\bar{u}s)\to J/\psi  p K^-$ at the quark level. We can also  search for the pentaquark states $P_{00,-}(4290)$,
$P_{00,+}(4410)$, $P_{01,-}(4300)$ and $P_{01,+}(4820)$   predicted in the present work in the decays $\Lambda_b^0\to J/\psi  p K^-$ as the same mechanism works.

\section*{Acknowledgements}
This  work is supported by National Natural Science Foundation,
Grant Numbers 11375063, 11235005, and Natural Science Foundation of Hebei province, Grant Number A2014502017.

\section*{Appendix}
The QCD spectral densities $\rho^1_{j_L j_H,i}(s)$ and $\widetilde{\rho}^0_{j_L j_H,i}(s)$ (with $i=0,\,3,\,4,\,5,\,6,\,8,\,9,\,10$) of the pentaquark states,

\begin{eqnarray}
\rho^1_{00,0}(s)&=&\frac{1}{491520\pi^8}\int dydz \, yz(1-y-z)^4\left(s-\overline{m}_c^2\right)^4\left(8s-3\overline{m}_c^2 \right) \, , \nonumber\\
\widetilde{\rho}^0_{00,0}(s)&=&\frac{1}{983040\pi^8}\int dydz \, (y+z)(1-y-z)^4\left(s-\overline{m}_c^2\right)^4\left(7s-2\overline{m}_c^2 \right) \, , \\
\rho^1_{01,0}(s)&=&\frac{1}{12280\pi^8}\int dydz \, yz(1-y-z)^4\left(s-\overline{m}_c^2\right)^4\left(8s-3\overline{m}_c^2 \right) \, , \nonumber\\
\widetilde{\rho}^0_{01,0}(s)&=&\frac{1}{491520\pi^8}\int dydz \, (y+z)(1-y-z)^4\left(s-\overline{m}_c^2\right)^4\left(7s-2\overline{m}_c^2 \right) \, ,
\end{eqnarray}

\begin{eqnarray}
\rho^1_{00,3}(s)&=&- \frac{m_c\langle \bar{q}q\rangle}{3072\pi^6}\int dydz \, (y+z)(1-y-z)^2\left(s-\overline{m}_c^2\right)^3  \, ,\nonumber \\
\widetilde{\rho}^0_{00,3}(s)&=&- \frac{m_c\langle \bar{q}q\rangle}{1536\pi^6}\int dydz \, (1-y-z)^2\left(s-\overline{m}_c^2\right)^3  \, ,\\
\rho^1_{01,3}(s)&=&- \frac{m_c\langle \bar{q}q\rangle}{1536\pi^6}\int dydz \, (y+z)(1-y-z)^2\left(s-\overline{m}_c^2\right)^3  \, ,\nonumber \\
\widetilde{\rho}^0_{01,3}(s)&=&- \frac{m_c\langle \bar{q}q\rangle}{384\pi^6}\int dydz \, (1-y-z)^2\left(s-\overline{m}_c^2\right)^3  \, ,
\end{eqnarray}

\begin{eqnarray}
\rho^1_{00,4}(s)&=&-\frac{m_c^2}{147456\pi^6} \langle\frac{\alpha_s GG}{\pi}\rangle\int dydz \left( \frac{z}{y^2}+\frac{y}{z^2}\right)(1-y-z)^4 \left(s-\overline{m}_c^2\right)\left(5s-3\overline{m}_c^2\right)  \nonumber\\
&&+\frac{19}{786432\pi^6} \langle\frac{\alpha_s GG}{\pi}\rangle\int dydz \left( y+z\right)(1-y-z)^3 \left(s-\overline{m}_c^2\right)^2\left(2s-\overline{m}_c^2\right)  \nonumber\\
&&+\frac{13}{131072\pi^6} \langle\frac{\alpha_s GG}{\pi}\rangle\int dydz \, yz(1-y-z)^2 \left(s-\overline{m}_c^2\right)^2\left(2s-\overline{m}_c^2\right) \, , \nonumber\\
\widetilde{\rho}^0_{00,4}(s)&=&-\frac{m_c^2}{147456\pi^6} \langle\frac{\alpha_s GG}{\pi}\rangle\int dydz \left( \frac{1}{y^2}+\frac{1}{z^2}+\frac{y}{z^3}+\frac{z}{y^3}\right)(1-y-z)^4 \left(s-\overline{m}_c^2\right)\left(2s-\overline{m}_c^2\right)  \nonumber\\
&&+\frac{1}{294912\pi^6} \langle\frac{\alpha_s GG}{\pi}\rangle\int dydz \left( \frac{y}{z^2}+\frac{z}{y^2}\right)(1-y-z)^4 \left(s-\overline{m}_c^2\right)^2\left(5s-2\overline{m}_c^2\right)  \nonumber\\
&&+\frac{19}{1179648\pi^6} \langle\frac{\alpha_s GG}{\pi}\rangle\int dydz \,(1-y-z)^3 \left(s-\overline{m}_c^2\right)^2\left(5s-2\overline{m}_c^2\right)  \nonumber\\
&&+\frac{13}{786432\pi^6} \langle\frac{\alpha_s GG}{\pi}\rangle\int dydz \,(y+z)(1-y-z)^2 \left(s-\overline{m}_c^2\right)^2\left(5s-2\overline{m}_c^2\right)  \, ,  \\
\rho^1_{01,4}(s)&=&-\frac{m_c^2}{36864\pi^6} \langle\frac{\alpha_s GG}{\pi}\rangle\int dydz \left( \frac{z}{y^2}+\frac{y}{z^2}\right)(1-y-z)^4 \left(s-\overline{m}_c^2\right)\left(5s-3\overline{m}_c^2\right)  \nonumber\\
&&+\frac{13}{32768\pi^6} \langle\frac{\alpha_s GG}{\pi}\rangle\int dydz \, yz(1-y-z)^2 \left(s-\overline{m}_c^2\right)^2\left(2s-\overline{m}_c^2\right) \, , \nonumber\\
\widetilde{\rho}^0_{01,4}(s)&=&-\frac{m_c^2}{73728\pi^6} \langle\frac{\alpha_s GG}{\pi}\rangle\int dydz \left( \frac{1}{y^2}+\frac{1}{z^2}+\frac{y}{z^3}+\frac{z}{y^3}\right)(1-y-z)^4 \left(s-\overline{m}_c^2\right)\left(2s-\overline{m}_c^2\right)  \nonumber\\
&&+\frac{1}{147456\pi^6} \langle\frac{\alpha_s GG}{\pi}\rangle\int dydz \left( \frac{y}{z^2}+\frac{z}{y^2}\right)(1-y-z)^4 \left(s-\overline{m}_c^2\right)^2\left(5s-2\overline{m}_c^2\right)  \nonumber\\
&&-\frac{19}{589824\pi^6} \langle\frac{\alpha_s GG}{\pi}\rangle\int dydz \,(1-y-z)^3 \left(s-\overline{m}_c^2\right)^2\left(5s-2\overline{m}_c^2\right)  \nonumber\\
&&+\frac{13}{393216\pi^6} \langle\frac{\alpha_s GG}{\pi}\rangle\int dydz \,(y+z)(1-y-z)^2 \left(s-\overline{m}_c^2\right)^2\left(5s-2\overline{m}_c^2\right)  \, ,
\end{eqnarray}

\begin{eqnarray}
\rho^1_{00,5}(s)&=& \frac{19m_c\langle \bar{q}g_s\sigma Gq\rangle}{32768\pi^6}\int dydz  \, (y+z)(1-y-z) \left(s-\overline{m}_c^2 \right)^2 \nonumber\\
&&-\frac{m_c\langle \bar{q}g_s\sigma Gq\rangle}{4096\pi^6}\int dydz  \, \left(\frac{y}{z}+\frac{z}{y}\right)(1-y-z)^2 \left(s-\overline{m}_c^2 \right)^2 \, ,\nonumber\\
\widetilde{\rho}^0_{00,5}(s)&=& \frac{19m_c\langle \bar{q}g_s\sigma Gq\rangle}{16384\pi^6}\int dydz  \, (1-y-z) \left(s-\overline{m}_c^2 \right)^2 \nonumber\\
&&-\frac{m_c\langle \bar{q}g_s\sigma Gq\rangle}{4096\pi^6}\int dydz  \, \left(\frac{1}{y}+\frac{1}{z}\right)(1-y-z)^2 \left(s-\overline{m}_c^2 \right)^2 \, ,\\
\rho^1_{01,5}(s)&=& \frac{19m_c\langle \bar{q}g_s\sigma Gq\rangle}{16384\pi^6}\int dydz  \, (y+z)(1-y-z) \left(s-\overline{m}_c^2 \right)^2 \nonumber\\
&&+\frac{m_c\langle \bar{q}g_s\sigma Gq\rangle}{32768\pi^6}\int dydz  \, \left(\frac{y}{z}+\frac{z}{y}\right)(1-y-z)^2 \left(s-\overline{m}_c^2 \right)^2 \nonumber\\
&&+\frac{m_c\langle \bar{q}g_s\sigma Gq\rangle}{49152\pi^6}\int dydz  \, \left(\frac{y}{z}+\frac{z}{y}\right)(1-y-z)^3 \left(s-\overline{m}_c^2 \right) \left(5s-3\overline{m}_c^2 \right)  \, ,\nonumber\\
\widetilde{\rho}^0_{01,5}(s)&=& \frac{7m_c\langle \bar{q}g_s\sigma Gq\rangle}{4096\pi^6}\int dydz  \, (1-y-z) \left(s-\overline{m}_c^2 \right)^2 \nonumber\\
&&+\frac{m_c\langle \bar{q}g_s\sigma Gq\rangle}{16384\pi^6}\int dydz  \, \left(\frac{1}{y}+\frac{1}{z}\right)(1-y-z)^2 \left(s-\overline{m}_c^2 \right)^2 \nonumber\\
&&-\frac{m_c\langle \bar{q}g_s\sigma Gq\rangle}{24576\pi^6}\int dydz  \, \left(\frac{1}{y}+\frac{1}{z}\right)(1-y-z)^3 \left(s-\overline{m}_c^2 \right)\left(2s-\overline{m}_c^2 \right)  \, ,
\end{eqnarray}

\begin{eqnarray}
\rho^1_{00,6}(s)&=&\frac{\langle\bar{q}q\rangle^2}{192\pi^4}\int dydz \,  yz(1-y-z)\left(s-\overline{m}_c^2 \right)\left(5s-3\overline{m}_c^2 \right)\,,\nonumber\\
\widetilde{\rho}^0_{00,6}(s)&=&\frac{\langle\bar{q}q\rangle^2}{192\pi^4}\int dydz \,  (y+z)(1-y-z)\left(s-\overline{m}_c^2 \right)\left(2s-\overline{m}_c^2 \right)\, , \\
\rho^1_{01,6}(s)&=&\frac{\langle\bar{q}q\rangle^2}{48\pi^4}\int dydz \,  yz(1-y-z)\left(s-\overline{m}_c^2 \right)\left(5s-3\overline{m}_c^2 \right)\,,\nonumber\\
\widetilde{\rho}^0_{01,6}(s)&=&\frac{\langle\bar{q}q\rangle^2}{96\pi^4}\int dydz \,  (y+z)(1-y-z)\left(s-\overline{m}_c^2 \right)\left(2s-\overline{m}_c^2 \right)\, ,
\end{eqnarray}

\begin{eqnarray}
\rho^1_{00,8}(s)&=&-\frac{35\langle\bar{q}q\rangle\langle\bar{q}g_s\sigma Gq\rangle}{6144\pi^4}\int dydz \,yz \left(4s-3\overline{m}_c^2\right)\nonumber\\
&&+\frac{\langle\bar{q}q\rangle\langle\bar{q}g_s\sigma Gq\rangle}{4096\pi^4}\int dydz \,(y+z)(1-y-z) \left(4s-3\overline{m}_c^2\right)\, ,\nonumber\\
\widetilde{\rho}^0_{00,8}(s)&=&-\frac{35\langle\bar{q}q\rangle\langle\bar{q}g_s\sigma Gq\rangle}{12288\pi^4}\int dydz \,(y+z) \left(3s-2\overline{m}_c^2\right)\nonumber\\
&&+\frac{\langle\bar{q}q\rangle\langle\bar{q}g_s\sigma Gq\rangle}{2048\pi^4}\int dydz \, (1-y-z) \left(3s-2\overline{m}_c^2\right)\, , \\
\rho^1_{01,8}(s)&=&-\frac{35\langle\bar{q}q\rangle\langle\bar{q}g_s\sigma Gq\rangle}{1536\pi^4}\int dydz \,yz \left(4s-3\overline{m}_c^2\right)\,,\nonumber\\
\widetilde{\rho}^0_{01,8}(s)&=&-\frac{35\langle\bar{q}q\rangle\langle\bar{q}g_s\sigma Gq\rangle}{6144\pi^4}\int dydz \,(y+z) \left(3s-2\overline{m}_c^2\right)\nonumber\\
&&-\frac{\langle\bar{q}q\rangle\langle\bar{q}g_s\sigma Gq\rangle}{1024\pi^4}\int dydz \, (1-y-z) \left(3s-2\overline{m}_c^2\right)\, ,
\end{eqnarray}

\begin{eqnarray}
\rho^1_{00,9}(s)&=&-\frac{m_c\langle\bar{q}q\rangle^3}{144\pi^2}\int_{y_i}^{y_f} dy   \, , \nonumber\\
\widetilde{\rho}^0_{00,9}(s)&=&-\frac{m_c\langle\bar{q}q\rangle^3}{72\pi^2}\int_{y_i}^{y_f} dy   \,, \\
\rho^1_{01,9}(s)&=&-\frac{m_c\langle\bar{q}q\rangle^3}{72\pi^2}\int_{y_i}^{y_f} dy   \, , \nonumber\\
\widetilde{\rho}^0_{01,9}(s)&=&-\frac{m_c\langle\bar{q}q\rangle^3}{18\pi^2}\int_{y_i}^{y_f} dy   \,,
\end{eqnarray}

\begin{eqnarray}
\rho^1_{00,10}(s)&=&\frac{19\langle\bar{q}g_s\sigma Gq\rangle^2}{24576\pi^4}\int_{y_i}^{y_f} dy \, y(1-y)\left[3+\widetilde{m}_c^2 \, \delta \left( s-\widetilde{m}_c^2\right)\right]\nonumber\\
&&-\frac{17\langle\bar{q}g_s\sigma Gq\rangle^2}{147456\pi^4}\int dydz \, (y+z)\left[3+\overline{m}_c^2 \, \delta \left( s-\overline{m}_c^2\right)\right] \, ,
\nonumber\\
\widetilde{\rho}^0_{00,10}(s)&=&\frac{19\langle\bar{q}g_s\sigma Gq\rangle^2}{49152\pi^4}\int_{y_i}^{y_f} dy \, \left[2+\widetilde{m}_c^2 \, \delta \left( s-\widetilde{m}_c^2\right)\right]\nonumber\\
&&-\frac{17\langle\bar{q}g_s\sigma Gq\rangle^2}{73728\pi^4}\int dydz \,\left[2+\overline{m}_c^2 \, \delta \left( s-\overline{m}_c^2\right)\right]\, , \\
\rho^1_{01,10}(s)&=&\frac{19\langle\bar{q}g_s\sigma Gq\rangle^2}{6144\pi^4}\int_{y_i}^{y_f} dy \, y(1-y)\left[3+\widetilde{m}_c^2 \, \delta \left( s-\widetilde{m}_c^2\right)\right] \, ,
\nonumber\\
\widetilde{\rho}^0_{01,10}(s)&=&\frac{19\langle\bar{q}g_s\sigma Gq\rangle^2}{24576\pi^4}\int_{y_i}^{y_f} dy \, \left[2+\widetilde{m}_c^2 \, \delta \left( s-\widetilde{m}_c^2\right)\right]\nonumber\\
&&+\frac{17\langle\bar{q}g_s\sigma Gq\rangle^2}{36864\pi^4}\int dydz \,\left[2+\overline{m}_c^2 \, \delta \left( s-\overline{m}_c^2\right)\right]\, ,
\end{eqnarray}
where $\int dydz=\int_{y_i}^{y_f}dy \int_{z_i}^{1-y}dz$, $y_{f}=\frac{1+\sqrt{1-4m_c^2/s}}{2}$,
$y_{i}=\frac{1-\sqrt{1-4m_c^2/s}}{2}$, $z_{i}=\frac{y
m_c^2}{y s -m_c^2}$, $\overline{m}_c^2=\frac{(y+z)m_c^2}{yz}$,
$ \widetilde{m}_c^2=\frac{m_c^2}{y(1-y)}$, $\int_{y_i}^{y_f}dy \to \int_{0}^{1}dy$, $\int_{z_i}^{1-y}dz \to \int_{0}^{1-y}dz$ when the $\delta$ functions $\delta\left(s-\overline{m}_c^2\right)$ and $\delta\left(s-\widetilde{m}_c^2\right)$ appear.

\end{document}